\documentclass[12pt]{report}
\usepackage{amssymb}
\usepackage{amsmath}

\usepackage{graphicx}
\begin{document}
\voffset = -2.9cm
\hoffset = -1.3cm
\def\itm{\newline \makebox[8mm]{}}
\def\ls{\makebox[8mm]{}}
\def\fra#1#2{\frac{#1}{#2}}
\def\fr#1#2{#1/#2}
\def\frl#1#2{\mbox{\large $\frac{#1}{\rule[-0mm]{0mm}{3.15mm} #2}$}}
\def\frn#1#2{\mbox{\normalsize $\frac{#1}{\rule[-0mm]{0mm}{3.15mm} #2}$}}
\def\frm#1#2{\mbox{\normalsize $\frac{#1}{\rule[-0mm]{0mm}{2.85mm} #2}$}}
\def\frn#1#2{\mbox{\normalsize $\frac{#1}{\rule[-0mm]{0mm}{3.15mm} #2}$}}
\def\hs#1{\mbox{\hspace{#1}}}
\def\b{\begin{equation}}
\def\e{\end{equation}}
\def\arccot{\mbox{arccot}}
\vspace*{6mm}
\makebox[\textwidth][c]
{\large \bf{FRW Universe Models in Conformally Flat Spacetime Coordinates}}
\vspace*{-1.0mm} \newline
\hspace*{11.4mm}
{\large \bf{III: Universe models with positive spatial curvature}}
\vspace{4mm} \newline
\makebox[\textwidth][c]
{\normalsize \O yvind Gr\o n$^{*}$ and Steinar Johannesen$^*$}
\vspace{1mm} \newline
\makebox[\textwidth][c]
{\scriptsize $*$ Oslo University College, Faculty of Engineering,
P.O.Box 4 St.Olavs Plass, N-0130 Oslo, Norway}
%
\vspace{3mm} \newline
{\bf \small Abstract}
{\small We deduce general expressions for the line element of universe
models with positive spatial curvature described by conformally flat
spacetime coordinates. Models with dust, radiation and vacuum energy
are exhibited. Discussing the existence of particle horizons we show
that there is continual annihilation of space, matter and energy in a
dust and radiation dominated universe, and continual creation in a
LIVE domined universe when conformal time is used in
Friedmann-Robertson-Walker models with positive spatial curvature.
A general procedure is given for finding coordinates to be used
in Penrose diagrams. We also calculate the age and the redshift
of some universe models using conformal time.}
%
%
\vspace{6mm} \newline
{\bf 1. Introduction.}
\vspace{3mm} \newline
In order to complete the description of the FRW universe models in
conformally flat spacetime (CFS) coordinates as given in part I and II
of this series [1,2], it remains to describe the universe models with
positive spatial curvature. From the general formalism presented in
part I of this series, we find in section 2 the coordinate
transformations to a CFS description of such universe models.
Combining the transformations described in part II of this series
with the inverse of the transformations in section 2, we obtain
Penrose diagrams in section 3. We also present some results concerning
physical properties of FRW universe models as described in
CFS coordinates, including the Hubble parameter, the CFS age of
the universe, the recession velocity and the cosmic redshift.
In this paper we shall refer to equations in reference [1] or [2]
by writing I or II:(equation number).
%
%
\vspace{5mm} \newline
{\bf 2. Universe models with positive spatial curvature}
\vspace{3mm} \newline
%
%
{\bf 2.1. Conformal coordinates in positively curved universe models}
\vspace{3mm} \newline
We shall now consider positively curved universe models.
Then we can introduce conformal coordinates $(T,R)$ by choosing
$a = 0$, $b = 0$, $c = 1$ and $d = 0$ in equation I:(33).
This gives the generating function
\begin{equation} \label{e_207}
f(x) = \tan (x / 2)
\mbox{ .}
\end{equation}
The transformation I:(28) then takes the form [3,4]
\begin{equation} \label{e_49}
T = \frac{\sin \eta}{\cos \eta + \cos \chi}
\mbox{\hspace{2mm} , \hspace{3mm}}
R = \frac{\sin \chi}{\cos \eta + \cos \chi}
\mbox{ ,}
\end{equation}
transforming the triangle defined by $0 < \chi < \pi$ and
$0 < \eta < \pi - \chi$ onto the first quadrant $R > 0$, $T > 0$, and
the triangle defined by $0 < \chi < \pi$ and $\pi + \chi < \eta < 2 \pi$
onto the fourth quadrant $R > 0$, $T < 0$. We also have a conformal
coordinate system $(\widetilde{R},\widetilde{T})$ given by the same
formulae \eqref{e_49}, which maps the triangle defined by $0 < \chi < \pi$
and $\pi - \chi < \eta < \pi + \chi$ onto the left half plane
$\widetilde{R} < 0$. Note that the negative sign of $\widetilde{R}$ can
be removed by changing the coordinates $\theta$ and $\phi$ corresponding
to a reflection through the origin in space. These domains are those
needed in universe models with dust and radiation. We will later specify
the domains needed for universe models with vacuum energy. The inverse
transformation is
\begin{equation} \label{e_51}
\cot \eta = \frl{1 -
\left(T^{2} - R^{2} \right)}{2 T}
\mbox{\hspace{2mm} , \hspace{3mm}}
\cot \chi = \frl{1 +
\left(T^{2} - R^{2} \right)}{2 R}
\mbox{ .}
\end{equation}
\itm The world lines of the cosmic reference particles,
$\chi = \chi_0 \,$, as given in the conformal system, are
\begin{equation} \label{e_10}
(R + a_4)^2 - T^2 = a_4^2 + 1
\mbox{ ,}
\end{equation}
where $a_4 = \cot \chi_0$, describing hyperbolae with centra at $(-a_4,0)$.
The simultaneity curves of the cosmic space, $\eta = \eta_0 \,$, are
\begin{equation} \label{e_11}
(T + b_4)^2 - R^2 = b_4^2 + 1
\mbox{ ,}
\end{equation}
where $b_4 = \cot \eta_0$, describing hyperbolae with centra at $(-b_4,0)$.
These curves are shown in Figure 1 in the first quadrant.
\vspace*{-2mm} \newline
\begin{picture}(50,192)(-96,-182)
\qbezier(198.8359, -51.8816)(195.8136, -55.1718)(193.0546, -58.2218)
\qbezier(193.0546, -58.2218)(190.2957, -61.2717)(187.7812, -64.1026)
\qbezier(187.7812, -64.1026)(185.2667, -66.9334)(182.9791, -69.5646)
\qbezier(182.9791, -69.5646)(180.6915, -72.1959)(178.6151, -74.6458)
\qbezier(178.6151, -74.6458)(176.5387, -77.0957)(174.6591, -79.3813)
\qbezier(174.6591, -79.3813)(172.7794, -81.6668)(171.0836, -83.8037)
\qbezier(171.0836, -83.8037)(169.3877, -85.9407)(167.8639, -87.9438)
\qbezier(167.8639, -87.9438)(166.3401, -89.9470)(164.9778, -91.8302)
\qbezier(164.9778, -91.8302)(163.6155, -93.7134)(162.4052, -95.4897)
\qbezier(162.4052, -95.4897)(161.1950, -97.2660)(160.1284, -98.9477)
\qbezier(160.1284, -98.9477)(159.0619, -100.6294)(158.1317, -102.2282)
\qbezier(158.1317, -102.2282)(157.2014, -103.8269)(156.4011, -105.3537)
\qbezier(156.4011, -105.3537)(155.6007, -106.8805)(154.9247, -108.3460)
\qbezier(154.9247, -108.3460)(154.2487, -109.8115)(153.6924, -111.2257)
\qbezier(153.6924, -111.2257)(153.1361, -112.6400)(152.6955, -114.0128)
\qbezier(152.6955, -114.0128)(152.2550, -115.3856)(151.9273, -116.7265)
\qbezier(151.9273, -116.7265)(151.5995, -118.0674)(151.3823, -119.3857)
\qbezier(151.3823, -119.3857)(151.1650, -120.7039)(151.0568, -122.0087)
\qbezier(151.0568, -122.0087)(150.9485, -123.3134)(150.9485, -124.6136)
\qbezier(180.8537, -42.0300)(178.1816, -45.2015)(175.7093, -48.2064)
\qbezier(175.7093, -48.2064)(173.2369, -51.2114)(170.9541, -54.0625)
\qbezier(170.9541, -54.0625)(168.6713, -56.9135)(166.5684, -59.6227)
\qbezier(166.5684, -59.6227)(164.4654, -62.3319)(162.5336, -64.9105)
\qbezier(162.5336, -64.9105)(160.6018, -67.4892)(158.8330, -69.9480)
\qbezier(158.8330, -69.9480)(157.0642, -72.4069)(155.4509, -74.7564)
\qbezier(155.4509, -74.7564)(153.8377, -77.1058)(152.3733, -79.3557)
\qbezier(152.3733, -79.3557)(150.9088, -81.6056)(149.5871, -83.7653)
\qbezier(149.5871, -83.7653)(148.2654, -85.9250)(147.0808, -88.0036)
\qbezier(147.0808, -88.0036)(145.8962, -90.0822)(144.8438, -92.0885)
\qbezier(144.8438, -92.0885)(143.7913, -94.0947)(142.8667, -96.0370)
\qbezier(142.8667, -96.0370)(141.9420, -97.9792)(141.1413, -99.8656)
\qbezier(141.1413, -99.8656)(140.3405, -101.7521)(139.6602, -103.5906)
\qbezier(139.6602, -103.5906)(138.9800, -105.4291)(138.4174, -107.2274)
\qbezier(138.4174, -107.2274)(137.8548, -109.0257)(137.4076, -110.7913)
\qbezier(137.4076, -110.7913)(136.9603, -112.5569)(136.6265, -114.2973)
\qbezier(136.6265, -114.2973)(136.2927, -116.0376)(136.0710, -117.7600)
\qbezier(136.0710, -117.7600)(135.8492, -119.4824)(135.7386, -121.1940)
\qbezier(135.7386, -121.1940)(135.6279, -122.9056)(135.6279, -124.6136)
\qbezier(157.4655, -33.9283)(155.9383, -36.6317)(154.5013, -39.2853)
\qbezier(154.5013, -39.2853)(153.0643, -41.9388)(151.7159, -44.5455)
\qbezier(151.7159, -44.5455)(150.3675, -47.1521)(149.1062, -49.7148)
\qbezier(149.1062, -49.7148)(147.8449, -52.2775)(146.6692, -54.7992)
\qbezier(146.6692, -54.7992)(145.4935, -57.3208)(144.4020, -59.8043)
\qbezier(144.4020, -59.8043)(143.3106, -62.2878)(142.3022, -64.7359)
\qbezier(142.3022, -64.7359)(141.2939, -67.1840)(140.3674, -69.5995)
\qbezier(140.3674, -69.5995)(139.4409, -72.0150)(138.5953, -74.4007)
\qbezier(138.5953, -74.4007)(137.7497, -76.7864)(136.9840, -79.1450)
\qbezier(136.9840, -79.1450)(136.2183, -81.5035)(135.5316, -83.8376)
\qbezier(135.5316, -83.8376)(134.8449, -86.1717)(134.2365, -88.4840)
\qbezier(134.2365, -88.4840)(133.6280, -90.7963)(133.0971, -93.0894)
\qbezier(133.0971, -93.0894)(132.5662, -95.3824)(132.1123, -97.6590)
\qbezier(132.1123, -97.6590)(131.6584, -99.9355)(131.2809, -102.1980)
\qbezier(131.2809, -102.1980)(130.9034, -104.4605)(130.6019, -106.7116)
\qbezier(130.6019, -106.7116)(130.3004, -108.9626)(130.0746, -111.2048)
\qbezier(130.0746, -111.2048)(129.8488, -113.4470)(129.6984, -115.6829)
\qbezier(129.6984, -115.6829)(129.5480, -117.9187)(129.4729, -120.1508)
\qbezier(129.4729, -120.1508)(129.3977, -122.3828)(129.3977, -124.6136)
\qbezier(125.7955, -117.6814)(127.6730, -117.6814)(129.5532, -117.5825)
\qbezier(129.5532, -117.5825)(131.4334, -117.4836)(133.3213, -117.2856)
\qbezier(133.3213, -117.2856)(135.2093, -117.0876)(137.1103, -116.7899)
\qbezier(137.1103, -116.7899)(139.0114, -116.4922)(140.9307, -116.0940)
\qbezier(140.9307, -116.0940)(142.8500, -115.6958)(144.7930, -115.1960)
\qbezier(144.7930, -115.1960)(146.7359, -114.6962)(148.7079, -114.0934)
\qbezier(148.7079, -114.0934)(150.6799, -113.4906)(152.6863, -112.7831)
\qbezier(152.6863, -112.7831)(154.6928, -112.0756)(156.7392, -111.2615)
\qbezier(156.7392, -111.2615)(158.7857, -110.4473)(160.8779, -109.5243)
\qbezier(160.8779, -109.5243)(162.9701, -108.6013)(165.1138, -107.5669)
\qbezier(165.1138, -107.5669)(167.2574, -106.5324)(169.4585, -105.3836)
\qbezier(169.4585, -105.3836)(171.6597, -104.2349)(173.9243, -102.9686)
\qbezier(173.9243, -102.9686)(176.1890, -101.7024)(178.5235, -100.3151)
\qbezier(178.5235, -100.3151)(180.8579, -98.9278)(183.2687, -97.4157)
\qbezier(183.2687, -97.4157)(185.6794, -95.9036)(188.1731, -94.2625)
\qbezier(188.1731, -94.2625)(190.6669, -92.6214)(193.2504, -90.8466)
\qbezier(193.2504, -90.8466)(195.8340, -89.0719)(198.5146, -87.1587)
\qbezier(198.5146, -87.1587)(201.1952, -85.2455)(203.9803, -83.1885)
\qbezier(203.9803, -83.1885)(206.7654, -81.1314)(209.6626, -78.9249)
\qbezier(125.7955, -94.5393)(127.0626, -94.5393)(128.3341, -94.4345)
\qbezier(128.3341, -94.4345)(129.6056, -94.3297)(130.8902, -94.1193)
\qbezier(130.8902, -94.1193)(132.1747, -93.9089)(133.4810, -93.5916)
\qbezier(133.4810, -93.5916)(134.7873, -93.2742)(136.1243, -92.8477)
\qbezier(136.1243, -92.8477)(137.4614, -92.4212)(138.8382, -91.8826)
\qbezier(138.8382, -91.8826)(140.2151, -91.3441)(141.6412, -90.6898)
\qbezier(141.6412, -90.6898)(143.0674, -90.0355)(144.5525, -89.2609)
\qbezier(144.5525, -89.2609)(146.0376, -88.4864)(147.5919, -87.5864)
\qbezier(147.5919, -87.5864)(149.1462, -86.6864)(150.7802, -85.6547)
\qbezier(150.7802, -85.6547)(152.4143, -84.6230)(154.1393, -83.4526)
\qbezier(154.1393, -83.4526)(155.8642, -82.2822)(157.6919, -80.9652)
\qbezier(157.6919, -80.9652)(159.5196, -79.6481)(161.4625, -78.1753)
\qbezier(161.4625, -78.1753)(163.4054, -76.7025)(165.4767, -75.0640)
\qbezier(165.4767, -75.0640)(167.5481, -73.4255)(169.7621, -71.6100)
\qbezier(169.7621, -71.6100)(171.9760, -69.7945)(174.3478, -67.7897)
\qbezier(174.3478, -67.7897)(176.7195, -65.7848)(179.2652, -63.5770)
\qbezier(179.2652, -63.5770)(181.8109, -61.3691)(184.5479, -58.9430)
\qbezier(184.5479, -58.9430)(187.2849, -56.5170)(190.2320, -53.8563)
\qbezier(190.2320, -53.8563)(193.1791, -51.1956)(196.3563, -48.2819)
\put(125.7955, -60.1818){\line(1, -1){ 64.4318}}
\put( 88.9773, -124.6136){\vector(1, 0){128.8636}}
\put(125.7955, -161.4318){\vector(0, 1){144.2045}}
\put(227.0455, -129.2159){\makebox(0,0)[]{\footnotesize{$R$}}}
\put(116.5909, -15.6932){\makebox(0,0)[]{\footnotesize{$T$}}}
\put(165.6818, -24.8977){\makebox(0,0)[]{\footnotesize{$\chi = \mbox{const}$}}}
\put(223.9773, -70.9205){\makebox(0,0)[]{\footnotesize{$\eta = \mbox{const}$}}}
\put(116.5909, -60.1818){\makebox(0,0)[]{\footnotesize{$P$}}}
\put(191.7614, -133.8182){\makebox(0,0)[]{\footnotesize{$H$}}}
\end{picture}
\vspace{-2mm} \newline
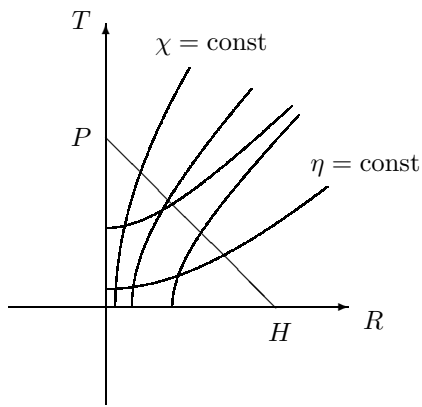
{\footnotesize \sf Figure 1. The first quadrant of the $(T,R)$-diagram
for universe models with positive spatial curvature with reference to
the conformal coordinate system defined in equation \eqref{e_49}.}
\vspace{0mm} \newline
\itm Note that all points on a simultaneity curve come arbitrary close
to the $R$-axis in the limit ${\eta}_0 \rightarrow 0$. Hence, in this
coordinate system, the Big Bang occurred everywhere at the moment $T = 0$.
Therefore there is no continual creation of conformal space in this
coordinate system. This is different from Big Bang as described with
reference to a conformal coordinate system of type I in a negatively
curved universe model where the Big Bang occurred along the light cone
in Figure 3 of reference [2].
\itm The line element now takes the conformally flat form [3 - 5]
\begin{equation} \label{e_53}
ds^{2} = \frac{4 \hs{0.9mm} a(\eta (T,R))^2}
{4 T^{2} + [1 - (T^{2} - R^{2})]^{2}} ds_{M}^{2}
\mbox{ ,}
\end{equation}
where the scale factor depends upon the matter and energy contents of
the universe. In a universe with dust and radiation the line
element is
\begin{equation} \label{e_22}
ds^2 = \left[ \frac{2 \alpha \{ \sqrt{4 T^2 + [1 - (T^2 - R^2)]^2}
- [1 - (T^2 - R^2)] \} + 4 \beta T}{4 T^2 + [1 - (T^2 - R^2)]^2}
\right]^{\hs{-0.0mm} 2} ds_M^2
\mbox{ .}
\end{equation}
\itm From equation I:(46) we obtain the following expression
for the recession velocity
\begin{equation} \label{e_104}
V = \frl{2TR}{1 + T^2 + R^2}
= \frl{\sin \eta \sin \chi}{1 + \cos \eta \cos \chi}
\mbox{ .}
\end{equation}
Once more the initial recession velocity vanishes.
%
%
%
\vspace{5mm} \newline
{\bf 2.2. Particle horizon in positively curved universe models with
radiation and dust}
\vspace{3mm} \newline
We shall here give a geometrical discussion of the particle horizon in
universe models with positive spatial curvature with reference to the
conformal coordinate system defined in equation \eqref{e_49}.
\itm In general, using cosmic coordinates, the particle horizon is
given in equation I:(9). Let us see how this can be deduced by
a geometrical consideration based upon Figure 1. In this figure
$P$ is an observation event at the point of time $T_0$.
\itm The horizon is the intersection of his backwards light cone with
the space defined by $\eta = 0$. This intersection is at the point $H$
in Figure 1, and from equation \eqref{e_49} it follows that the horizon
is represented by the $R$-axis in the figure.
\itm The conformal coordinates of the point $H$ are $T_H = 0$ and
$R_H = T_0$. Inserting this into equation \eqref{e_10} of a hyperbola
$\chi = {\chi}_H$ gives
\begin{equation} \label{e_28}
\cot {\chi}_H = \frac{1 - T_0^2}{2 T_0}
\mbox{ .}
\end{equation}
We also need to find the $\eta$-coordinate of $P$. The conformal
coordinates of $P$ are $T_P = T_0$ and $R_P = 0$. Inserting this into
equation \eqref{e_11} of a hyperbola $\eta = {\eta}_P$ gives
\begin{equation} \label{e_29}
\cot {\eta}_P = \frac{1 - T_0^2}{2 T_0}
\mbox{ .}
\end{equation}
Hence ${\chi}_H = {\eta}_P$ in accordance with equation I:(10).
\itm In a positively curved universe with radiation and dust
\begin{equation} \label{e_16}
a = \alpha (1 - \cos \eta) + \beta \sin \eta
\end{equation}
and
\begin{equation} \label{e_12}
t = \alpha (\eta - \sin \eta)
+ \beta (1 - \cos \eta)
\mbox{ ,}
\end{equation}
where
\begin{equation} \label{e_278}
\alpha = \frl{{\Omega}_{m0}}{2({\Omega}_{0} - 1)}
\mbox{\hspace{2mm} and \hspace{2mm}}
\beta = \sqrt{\frl{{\Omega}_{\gamma 0}}{{\Omega}_{0} - 1}}
\mbox{ .}
\end{equation}
For this universe model the relationship between the cosmic time and
conformal time at $R = 0$ is
\begin{equation} \label{e_163}
t = \alpha \left( \arcsin \frl{2T}{1 + T^2} - \frl{2T}{1 + T^2} \right)
+ \beta \hs{0.5mm} \frl{2T^2}{1 + T^2}
\mbox{ .}
\end{equation}
The scale factor $a(\eta)$ increases from zero at ${\eta}_1 = 0$ to a
maximal value at
\begin{equation} \label{e_144}
{\eta}_2 = \pi - \arctan \frl{\beta}{\alpha}
\mbox{ ,}
\end{equation}
and decreases to zero at ${\eta}_3 = 2 {\eta}_2$. In a radiation
dominated universe $\alpha = 0$ which gives ${\eta}_3 = \pi$. In a
dust dominated universe $\beta = 0$ giving ${\eta}_3 = 2 \pi$.
\itm In a closed universe the coordinate $\chi$ is defined in the interval
$<0 ,\pi>$ which means that the whole universe is covered except some
points where the coordinates are not uniquely defined. Imagine an
observer at $\chi = 0$. At a point of time $\eta$ he can see objects at
$\chi < {\chi}_H$ where ${\chi}_H = \eta$ is the radius of his particle
horizon. When ${\chi}_H = \pi$ the whole universe is inside the horizon.
Then the observer can see all of the universe. In a universe existing at
$\eta > \pi$ there is no longer any particle horizon. In a universe with
$k = 1$ and radiation only the Big Crunch occurs at $\eta = \pi$,
corresponding to $T = 0$.
The whole univese can only be seen at this final moment. In a matter
dominated universe the Big Crunch occurs at $\eta = 2 \pi$. Then the
whole universe can be seen at the moment of maximum extension. At this
moment the particle horizon vanishes.
\itm Consider a free particle with $\chi = {\chi}_0$ in a universe
with dust. When the parametric time approaches $\pi - {\chi}_0$,
then $T \rightarrow \infty$ and $R \rightarrow \infty$.
The conformal clocks go increasingly fast relative to the cosmic and
the parametric clocks. The relative rate approaches infinity as
$\eta \rightarrow \pi - {\chi}_0$. Hence the spacetime can not be covered
by a single conformal coordinate system. We must use two coordinate systems
$(T,R)$ and $(\widetilde{T},\widetilde{R})$ covering the domains I and II
in the $(\eta,\chi)$-plane as shown in Figure 2.
\vspace{0mm} \newline
\vspace*{3mm} \newline
\begin{picture}(50,192)(14,-182)
\put(125.7955, -69.3864){\line(1, -1){ 55.2273}}
\put(125.7955, -69.3864){\line(1, 1){ 55.2273}}
\put(125.7955, -14.1591){\line(1, 0){ 55.2273}}
\put(181.0227, -124.6136){\line(0, 1){110.4545}}
\put(164.1477, -124.6136){\line(0, 1){110.4545}}
\put( 95.1136, -124.6136){\vector(1, 0){116.5909}}
\put(125.7955, -155.2955){\vector(0, 1){168.7500}}
\put(220.9091, -129.2159){\makebox(0,0)[]{\footnotesize{$\chi$}}}
\put(116.5909,  14.9886){\makebox(0,0)[]{\footnotesize{$\eta$}}}
\put(116.8977, -14.1591){\makebox(0,0)[]{\footnotesize{$2 \pi$}}}
\put(118.1250, -69.3864){\makebox(0,0)[]{\footnotesize{$\pi$}}}
\put(181.9432, -133.8182){\makebox(0,0)[]{\footnotesize{$\pi$}}}
\put(164.7614, -133.8182){\makebox(0,0)[]{\footnotesize{$\chi_0$}}}
\put(170.8977, -103.1364){\makebox(0,0)[]{\footnotesize{$B$}}}
\put(144.2045, -106.2045){\makebox(0,0)[]{\small{$I$}}}
\put(144.2045, -32.5682){\makebox(0,0)[]{\small{$I$}}}
\put(153.4091, -69.3864){\makebox(0,0)[]{\small{$II$}}}
\end{picture}
\begin{picture}(50,192)(-146,-182)
\qbezier(175.7547, -93.0451)(175.7547, -93.4653)(175.0947, -93.8799)
\qbezier(175.0947, -92.2103)(174.4346, -91.7958)(173.1325, -91.3983)
\qbezier(175.0947, -93.8799)(174.4346, -94.2944)(173.1325, -94.6919)
\qbezier(173.1325, -94.6919)(171.8304, -95.0894)(169.9217, -95.4590)
\qbezier(169.9217, -90.6312)(168.0130, -90.2616)(165.5499, -89.9299)
\qbezier(169.9217, -95.4590)(168.0130, -95.8286)(165.5499, -96.1602)
\qbezier(165.5499, -96.1602)(163.0868, -96.4919)(160.1364, -96.7765)
\qbezier(160.1364, -89.3137)(157.1859, -89.0290)(153.8287, -88.7992)
\qbezier(160.1364, -96.7765)(157.1859, -97.0611)(153.8287, -97.2910)
\qbezier(153.8287, -97.2910)(150.4715, -97.5208)(146.7990, -97.6897)
\qbezier(146.7990, -88.4005)(143.1266, -88.2317)(139.2391, -88.1285)
\qbezier(146.7990, -97.6897)(143.1266, -97.8585)(139.2391, -97.9616)
\qbezier(139.2391, -97.9616)(135.3515, -98.0648)(131.3550, -98.0995)
\qbezier(131.3550, -87.9906)(127.3585, -87.9559)(123.3620, -87.9906)
\qbezier(131.3550, -98.0995)(127.3585, -98.1342)(123.3620, -98.0995)
\qbezier(123.3620, -98.0995)(119.3654, -98.0648)(115.4779, -97.9616)
\qbezier(115.4779, -88.1285)(111.5904, -88.2317)(107.9179, -88.4005)
\qbezier(115.4779, -97.9616)(111.5904, -97.8585)(107.9179, -97.6897)
\qbezier(107.9179, -97.6897)(104.2455, -97.5208)(100.8883, -97.2910)
\qbezier(100.8883, -88.7992)( 97.5310, -89.0290)( 94.5806, -89.3137)
\qbezier(100.8883, -97.2910)( 97.5310, -97.0611)( 94.5806, -96.7765)
\qbezier( 94.5806, -96.7765)( 91.6302, -96.4919)( 89.1671, -96.1602)
\qbezier( 89.1671, -89.9299)( 86.7039, -90.2616)( 84.7953, -90.6312)
\qbezier( 89.1671, -96.1602)( 86.7039, -95.8286)( 84.7953, -95.4590)
\qbezier( 84.7953, -95.4590)( 82.8866, -95.0894)( 81.5845, -94.6919)
\qbezier( 81.5845, -91.3983)( 80.2824, -91.7958)( 79.6223, -92.2103)
\qbezier( 81.5845, -94.6919)( 80.2824, -94.2944)( 79.6223, -93.8799)
\qbezier( 79.6223, -93.8799)( 78.9623, -93.4653)( 78.9623, -93.0451)
\qbezier(127.3585, -69.1697)(129.9057, -70.4263)(132.4528, -71.6829)
\qbezier(127.3585, -69.1697)(127.3585, -66.2194)(127.3585, -63.2690)
\qbezier(137.5472, -74.1961)(140.0943, -75.4527)(142.6415, -76.7093)
\qbezier(127.3585, -57.3684)(127.3585, -54.4180)(127.3585, -51.4677)
\qbezier(147.7358, -79.2225)(150.2830, -80.4791)(152.8302, -81.7357)
\qbezier(127.3585, -45.5670)(127.3585, -42.6167)(127.3585, -39.6663)
\qbezier(157.9245, -84.2489)(160.4717, -85.5055)(163.0189, -86.7621)
\qbezier(127.3585, -33.7657)(127.3585, -30.8153)(127.3585, -27.8650)
\qbezier(168.1132, -89.2753)(170.6604, -90.5319)(173.2075, -91.7885)
\qbezier(127.3585, -21.9643)(127.3585, -19.0140)(127.3585, -16.0637)
\qbezier(134.2722, -72.5805)(134.4446, -72.1967)(134.5774, -71.7960)
\qbezier(134.5774, -71.7960)(134.7101, -71.3953)(134.8017, -70.9822)
\qbezier(134.8017, -70.9822)(134.8933, -70.5691)(134.9427, -70.1481)
\qbezier(134.9427, -70.1481)(134.9922, -69.7271)(134.9989, -69.3031)
\qbezier(134.9989, -69.3031)(135.0057, -68.8790)(134.9696, -68.4565)
\qbezier(134.9696, -68.4565)(134.9336, -68.0341)(134.8552, -67.6180)
\qbezier(134.8552, -67.6180)(134.7767, -67.2019)(134.6568, -66.7968)
\qbezier(134.6568, -66.7968)(134.5369, -66.3917)(134.3768, -66.0021)
\qbezier(134.3768, -66.0021)(134.2167, -65.6126)(134.0182, -65.2429)
\qbezier(134.0182, -65.2429)(133.8198, -64.8733)(133.5852, -64.5276)
\qbezier(133.5852, -64.5276)(133.3506, -64.1820)(133.0824, -63.8643)
\qbezier(133.0824, -63.8643)(132.8143, -63.5466)(132.5156, -63.2604)
\qbezier(132.5156, -63.2604)(132.2170, -62.9741)(131.8911, -62.7225)
\qbezier(131.8911, -62.7225)(131.5653, -62.4710)(131.2159, -62.2568)
\qbezier(131.2159, -62.2568)(130.8666, -62.0427)(130.4975, -61.8685)
\qbezier(130.4975, -61.8685)(130.1285, -61.6943)(129.7440, -61.5619)
\qbezier(129.7440, -61.5619)(129.3595, -61.4295)(128.9638, -61.3403)
\qbezier(128.9638, -61.3403)(128.5681, -61.2512)(128.1657, -61.2064)
\qbezier(128.1657, -61.2064)(127.7632, -61.1616)(127.3585, -61.1616)
\qbezier(180.8491, -69.1697)(180.8491, -64.5247)(180.1195, -59.9431)
\qbezier(180.8491, -69.1697)(180.8491, -73.8146)(180.1195, -78.3963)
\qbezier(180.1195, -59.9431)(179.3900, -55.3615)(177.9508, -50.9682)
\qbezier(180.1195, -78.3963)(179.3900, -82.9779)(177.9508, -87.3711)
\qbezier(177.9508, -50.9682)(176.5116, -46.5749)(174.4020, -42.4898)
\qbezier(177.9508, -87.3711)(176.5116, -91.7644)(174.4020, -95.8496)
\qbezier(174.4020, -42.4898)(172.2925, -38.4047)(169.5701, -34.7392)
\qbezier(174.4020, -95.8496)(172.2925, -99.9347)(169.5701, -103.6002)
\qbezier(169.5701, -34.7392)(166.8477, -31.0736)(163.5867, -27.9277)
\qbezier(169.5701, -103.6002)(166.8477, -107.2657)(163.5867, -110.4117)
\qbezier(163.5867, -27.9277)(160.3257, -24.7818)(156.6151, -22.2412)
\qbezier(163.5867, -110.4117)(160.3257, -113.5576)(156.6151, -116.0982)
\qbezier(156.6151, -22.2412)(152.9044, -19.7006)(148.8454, -17.8348)
\qbezier(156.6151, -116.0982)(152.9044, -118.6387)(148.8454, -120.5046)
\qbezier(148.8454, -17.8348)(144.7864, -15.9689)(140.4896, -14.8287)
\qbezier(148.8454, -120.5046)(144.7864, -122.3704)(140.4896, -123.5107)
\qbezier(140.4896, -14.8287)(136.1929, -13.6884)(131.7757, -13.3048)
\qbezier(140.4896, -123.5107)(136.1929, -124.6510)(131.7757, -125.0346)
\qbezier(131.7757, -13.3048)(127.3585, -12.9212)(122.9413, -13.3048)
\qbezier(131.7757, -125.0346)(127.3585, -125.4181)(122.9413, -125.0346)
\qbezier(122.9413, -13.3048)(118.5241, -13.6884)(114.2273, -14.8287)
\qbezier(122.9413, -125.0346)(118.5241, -124.6510)(114.2273, -123.5107)
\qbezier(114.2273, -14.8287)(109.9306, -15.9689)(105.8716, -17.8348)
\qbezier(114.2273, -123.5107)(109.9306, -122.3704)(105.8716, -120.5046)
\qbezier(105.8716, -17.8348)(101.8125, -19.7006)( 98.1019, -22.2412)
\qbezier(105.8716, -120.5046)(101.8125, -118.6387)( 98.1019, -116.0982)
\qbezier( 98.1019, -22.2412)( 94.3913, -24.7818)( 91.1303, -27.9277)
\qbezier( 98.1019, -116.0982)( 94.3913, -113.5576)( 91.1303, -110.4117)
\qbezier( 91.1303, -27.9277)( 87.8693, -31.0736)( 85.1469, -34.7392)
\qbezier( 91.1303, -110.4117)( 87.8693, -107.2657)( 85.1469, -103.6002)
\qbezier( 85.1469, -34.7392)( 82.4245, -38.4047)( 80.3149, -42.4898)
\qbezier( 85.1469, -103.6002)( 82.4245, -99.9347)( 80.3149, -95.8496)
\qbezier( 80.3149, -42.4898)( 78.2054, -46.5749)( 76.7662, -50.9682)
\qbezier( 80.3149, -95.8496)( 78.2054, -91.7644)( 76.7662, -87.3711)
\qbezier( 76.7662, -50.9682)( 75.3270, -55.3615)( 74.5975, -59.9431)
\qbezier( 76.7662, -87.3711)( 75.3270, -82.9779)( 74.5975, -78.3963)
\qbezier( 74.5975, -59.9431)( 73.8679, -64.5247)( 73.8679, -69.1697)
\qbezier( 74.5975, -78.3963)( 73.8679, -73.8146)( 73.8679, -69.1697)
\put(143.9151, -63.8310){\makebox(0,0)[]{\footnotesize{$\chi_0$}}}
\put(126.0849, -113.2140){\makebox(0,0)[]{\small{$II$}}}
\put(113.3491, -59.8270){\makebox(0,0)[]{\small{$I$}}}
\end{picture}
\vspace{-6mm} \newline
{\footnotesize \sf Figure 2. The domains of the conformal coordinate
systems defined in equation \eqref{e_49} for dust dominated universe
models. The right hand part of the figure shows the spherical $3$-space
of the closed universe model, suppressing the angular coordinate $\theta$.
The coordinate $\chi$ has the value $\chi = 0$ at the top of the figure
which we call the North pole, and the value $\chi = \pi$ at the South pole.
At $\chi = {\chi}_0$ the point $B$ in the left hand part of the figure
corresponds to the circle between the regions I and II on the right hand
part.}
\vspace{0mm} \newline
\itm The particle with $\chi = {\chi}_0$ enters the domain II at the
parametric time $\pi - {\chi}_0$, and comes back to domain I at the
time $\pi + {\chi}_0$.
The motion of the particle in the conformal coordinate systems looks
very strange. At the conformal time $T = 0$ the particle starts moving
upwards from the $R$-axis in Figure 3 along a hyperbola in the first
quadrant. As the parametric time approches $\pi - {\chi}_0$, the
conformal time and radius approaches infinity. The further motion of
the particle must be described in the other conformal coordinate system.
At the parametric time $\pi + {\chi}_0$ the particle enters the original
conformal coordinate system at infinite past and from infinitely far
away, arriving at the point of departure at a parametric time $2 \pi$
corresponding to a conformal time $T = 0$.
\vspace{1mm} \newline
\begin{picture}(50,182)(64,-202)
\qbezier(188.2363, -67.5168)(186.2859, -69.6591)(184.5018, -71.6511)
\qbezier(173.4666, -81.0087)(177.9743, -79.0818)(177.9743, -79.0818)
\qbezier(176.8112, -83.8441)(177.9743, -79.0818)(177.9743, -79.0818)
\qbezier(184.5018, -71.6511)(182.7178, -73.6431)(181.0886, -75.4978)
\qbezier(181.0886, -75.4978)(179.4593, -77.3524)(177.9743, -79.0818)
\qbezier(177.9743, -79.0818)(176.4893, -80.8111)(175.1389, -82.4264)
\qbezier(175.1389, -82.4264)(173.7885, -84.0416)(172.5639, -85.5533)
\qbezier(172.5639, -85.5533)(171.3393, -87.0649)(170.2326, -88.4827)
\qbezier(170.2326, -88.4827)(169.1259, -89.9005)(168.1299, -91.2337)
\qbezier(168.1299, -91.2337)(167.1339, -92.5669)(166.2421, -93.8241)
\qbezier(166.2421, -93.8241)(165.3503, -95.0813)(164.5570, -96.2708)
\qbezier(164.5570, -96.2708)(163.7637, -97.4602)(163.0637, -98.5895)
\qbezier(163.0637, -98.5895)(162.3637, -99.7188)(161.7525, -100.7954)
\qbezier(161.7525, -100.7954)(161.1413, -101.8719)(160.6149, -102.9027)
\qbezier(160.6149, -102.9027)(160.0884, -103.9335)(159.6434, -104.9252)
\qbezier(159.6434, -104.9252)(159.1984, -105.9169)(158.8318, -106.8759)
\qbezier(158.8318, -106.8759)(158.4653, -107.8349)(158.1749, -108.7675)
\qbezier(158.1749, -108.7675)(157.8845, -109.7001)(157.6683, -110.6123)
\qbezier(157.6683, -110.6123)(157.4521, -111.5245)(157.3087, -112.4222)
\qbezier(157.3087, -112.4222)(157.1654, -113.3200)(157.0940, -114.2091)
\qbezier(157.0940, -114.2091)(157.0225, -115.0982)(157.0225, -115.9844)
\put( 69.6094, -115.9844){\vector(1, 0){118.1250}}
\put(128.6719, -183.4844){\vector(0, 1){135.0000}}
\put(198.2812, -119.1484){\makebox(0,0)[]{\footnotesize{$R$}}}
\put(122.3437, -41.1016){\makebox(0,0)[]{\footnotesize{$T$}}}
\end{picture}
\begin{picture}(50,182)(-36,-202)
\qbezier( 81.7637, -67.5168)( 84.3184, -70.3228)( 86.5907, -72.8745)
\qbezier(100.0579, -84.6045)( 94.6741, -82.2030)( 94.6741, -82.2030)
\qbezier( 95.9832, -87.9509)( 94.6741, -82.2030)( 94.6741, -82.2030)
\qbezier( 86.5907, -72.8745)( 88.8631, -75.4261)( 90.8783, -77.7517)
\qbezier( 90.8783, -77.7517)( 92.8935, -80.0774)( 94.6741, -82.2030)
\qbezier( 94.6741, -82.2030)( 96.4547, -84.3285)( 98.0206, -86.2777)
\qbezier( 98.0206, -86.2777)( 99.5864, -88.2269)(100.9549, -90.0214)
\qbezier(100.9549, -90.0214)(102.3234, -91.8159)(103.5098, -93.4757)
\qbezier(103.5098, -93.4757)(104.6963, -95.1356)(105.7138, -96.6792)
\qbezier(105.7138, -96.6792)(106.7314, -98.2229)(107.5914, -99.6676)
\qbezier(107.5914, -99.6676)(108.4515, -101.1123)(109.1636, -102.4741)
\qbezier(109.1636, -102.4741)(109.8757, -103.8359)(110.4479, -105.1301)
\qbezier(110.4479, -105.1301)(111.0200, -106.4242)(111.4585, -107.6651)
\qbezier(111.4585, -107.6651)(111.8970, -108.9060)(112.2068, -110.1074)
\qbezier(112.2068, -110.1074)(112.5166, -111.3088)(112.7011, -112.4842)
\qbezier(112.7011, -112.4842)(112.8856, -113.6595)(112.9468, -114.8220)
\qbezier(112.9468, -114.8220)(113.0081, -115.9844)(112.9468, -117.1468)
\qbezier(112.9468, -117.1468)(112.8856, -118.3092)(112.7011, -119.4846)
\qbezier(112.7011, -119.4846)(112.5166, -120.6600)(112.2068, -121.8614)
\qbezier(112.2068, -121.8614)(111.8970, -123.0628)(111.4585, -124.3037)
\qbezier(111.4585, -124.3037)(111.0200, -125.5445)(110.4479, -126.8387)
\qbezier(110.4479, -126.8387)(109.8757, -128.1328)(109.1636, -129.4946)
\qbezier(109.1636, -129.4946)(108.4515, -130.8565)(107.5914, -132.3012)
\qbezier(107.5914, -132.3012)(106.7314, -133.7459)(105.7138, -135.2895)
\qbezier(105.7138, -135.2895)(104.6963, -136.8332)(103.5098, -138.4930)
\qbezier(103.5098, -138.4930)(102.3234, -140.1529)(100.9549, -141.9474)
\qbezier(100.9549, -141.9474)( 99.5864, -143.7419)( 98.0206, -145.6911)
\qbezier( 98.0206, -145.6911)( 96.4547, -147.6402)( 94.6741, -149.7658)
\qbezier( 94.6741, -149.7658)( 92.8935, -151.8914)( 90.8783, -154.2170)
\qbezier( 90.8783, -154.2170)( 88.8631, -156.5426)( 86.5907, -159.0943)
\qbezier( 86.5907, -159.0943)( 84.3184, -161.6459)( 81.7637, -164.4520)
\put( 69.6094, -115.9844){\vector(1, 0){118.1250}}
\put(128.6719, -183.4844){\vector(0, 1){135.0000}}
\put(198.2812, -119.1484){\makebox(0,0)[]{\footnotesize{$\widetilde{R}$}}}
\put(122.3437, -41.1016){\makebox(0,0)[]{\footnotesize{$\widetilde{T}$}}}
\end{picture}
\begin{picture}(50,182)(-136,-202)
\qbezier(157.0225, -115.9844)(157.0225, -116.8706)(157.0940, -117.7597)
\qbezier(160.9071, -134.6037)(160.6149, -129.0660)(160.6149, -129.0660)
\qbezier(165.2203, -132.1548)(160.6149, -129.0660)(160.6149, -129.0660)
\qbezier(157.0940, -117.7597)(157.1654, -118.6488)(157.3087, -119.5465)
\qbezier(157.3087, -119.5465)(157.4521, -120.4442)(157.6683, -121.3565)
\qbezier(157.6683, -121.3565)(157.8845, -122.2687)(158.1749, -123.2012)
\qbezier(158.1749, -123.2012)(158.4653, -124.1338)(158.8318, -125.0929)
\qbezier(158.8318, -125.0929)(159.1984, -126.0519)(159.6434, -127.0436)
\qbezier(159.6434, -127.0436)(160.0884, -128.0353)(160.6149, -129.0660)
\qbezier(160.6149, -129.0660)(161.1413, -130.0968)(161.7525, -131.1734)
\qbezier(161.7525, -131.1734)(162.3637, -132.2499)(163.0637, -133.3793)
\qbezier(163.0637, -133.3793)(163.7637, -134.5086)(164.5570, -135.6980)
\qbezier(164.5570, -135.6980)(165.3503, -136.8874)(166.2421, -138.1446)
\qbezier(166.2421, -138.1446)(167.1339, -139.4019)(168.1299, -140.7351)
\qbezier(168.1299, -140.7351)(169.1259, -142.0683)(170.2326, -143.4861)
\qbezier(170.2326, -143.4861)(171.3393, -144.9039)(172.5639, -146.4155)
\qbezier(172.5639, -146.4155)(173.7885, -147.9271)(175.1389, -149.5424)
\qbezier(175.1389, -149.5424)(176.4893, -151.1576)(177.9743, -152.8870)
\qbezier(177.9743, -152.8870)(179.4593, -154.6163)(181.0886, -156.4710)
\qbezier(181.0886, -156.4710)(182.7178, -158.3256)(184.5018, -160.3177)
\qbezier(184.5018, -160.3177)(186.2859, -162.3097)(188.2363, -164.4520)
\put( 69.6094, -115.9844){\vector(1, 0){118.1250}}
\put(128.6719, -183.4844){\vector(0, 1){135.0000}}
\put(198.2812, -119.1484){\makebox(0,0)[]{\footnotesize{$R$}}}
\put(122.3437, -41.1016){\makebox(0,0)[]{\footnotesize{$T$}}}
\end{picture}
%
\vspace*{-4mm} \newline
\hspace*{19.5mm} (a) \hspace{46mm} (b) \hspace{46mm} (c)
\vspace*{6mm} \newline
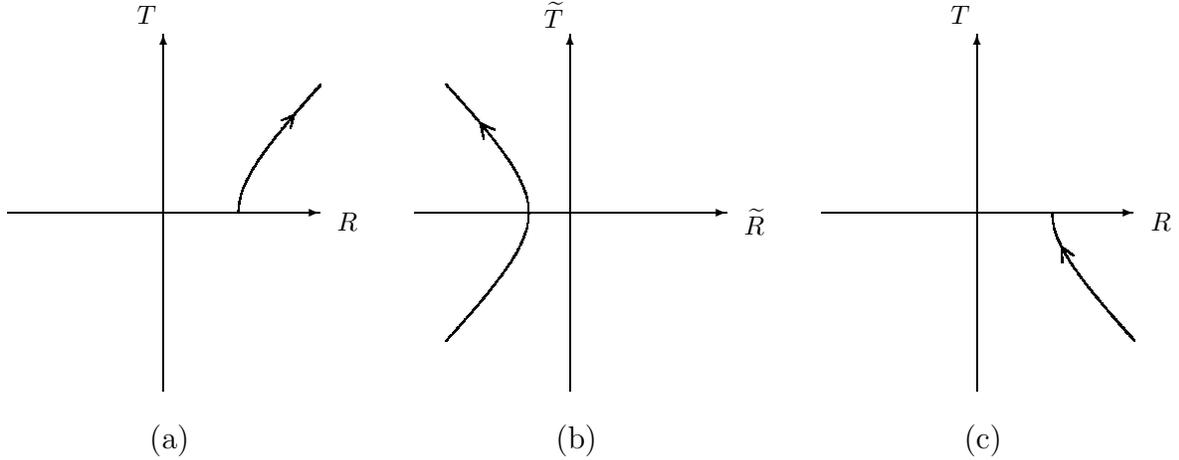
{\footnotesize \sf Figure 3. Minkowski diagram showing the motion of a
free particle in a dust dominated universe model with positive spatial
curvature, with reference to the conformal coordinate systems defined in
equation \eqref{e_49}. The world line of the particle in the
$(\eta,\chi)$-system is shown in Figure 2. (a) Imagine an observer
following the particle. Initially he observes clocks and meter sticks
showing $T$ and $R$. There is a Big Bang at $T = 0$. After the Big Bang
the observer finds that $T$ and $R$ increase towards infinity.
(b) The clocks and meter sticks are now replaced by new ones showing
$\widetilde{T}$ and $\widetilde{R}$. They have values increasing from
minus infinity. At $\widetilde{T} = 0$ the expansion stops, and the
universe starts contracting. Eventually $\widetilde{T}$ approaches
infinity and $\widetilde{R}$ minus infinity, corresponding to a
reflection through the origin. (c) Now the clocks
and meter sticks are replaced by the old ones, but adjusted so that
the time $T$ shown on the clocks increases from minus infinity and the
distance $R$ shown on the meter sticks decreases from infinity. At the
final moment, $T = 0$, there is a Big Crunch.
The condition that the particle shall enter the region I again is
that $\chi_0 < \pi - 2 \arctan (\beta / \alpha)$.}
\vspace{0mm} \newline
\itm In Figure 4 we have shown a backwards light cone of an observer at
the North pole at the parametric time
${\eta}_{\hs{0.3mm} 0} \in \hs{-1.3mm} < \hs{-0.5mm} \pi,2 \pi \hs{-0.5mm} >$.
$P$ is the observation event. The emission of the light is at
${\eta}_A = 0 \,$, ${\chi}_A = 2 \pi - {\eta}_{\hs{0.3mm} 0}$. At the
parametric time $({\eta}_{\hs{0.3mm} 0} - \pi) / 2$
the light passes from the region I to the region II, and at time
${\eta}_{\hs{0.3mm} 0} - \pi$ it passes the South pole on the sphere
in Figure 2. This looks like a reflection in Figure 4. Then, at the time
$({\eta}_{\hs{0.3mm} 0} + \pi) / 2$ it passes into the region I again,
arriving at $P$ at $\eta = {\eta}_{\hs{0.3mm} 0}$. The corresponding paths
in the conformal coordinate systems are shown in Figure 5. The conformal
radius of the emission point is
\begin{equation} \label{e_145}
R_A = - \frl{\sin {\eta}_{\hs{0.3mm} 0}}{1 + \cos {\eta}_{\hs{0.3mm} 0}}
\end{equation}
which is positive. If the light had been emitted from $A$ in the opposite
direction, it would have moved directly towards the North pole, not around
the universe via the South pole. This means that $A$ would have been seen
by the observer at the North pole much earlier than at ${\eta}_0$. Hence
for $\pi < {\eta}_0 < 2 \pi$, all of the universe could have been seen,
and there is no horizon. The closed universe model with radiation and dust
has a horizon only for ${\eta}_0 < \pi$. In a dust dominated universe the
horizon exists during the expansion era, and vanishes as the universe
reaches its maximum size.
\vspace{0mm} \newline
\vspace*{3mm} \newline
\begin{picture}(50,202)(-66,-172)
\put(125.7955, -69.3864){\line(1, -1){ 55.2273}}
\put(125.7955, -69.3864){\line(1, 1){ 55.2273}}
\put(125.7955, -14.1591){\line(1, 0){ 55.2273}}
\put(181.0227, -124.6136){\line(0, 1){110.4545}}
\put(125.7955, -41.7727){\line(1, -1){ 55.2273}}
\qbezier(158.9318, -71.2273)(153.4091, -69.3864)(153.4091, -69.3864)
\qbezier(155.2500, -74.9091)(153.4091, -69.3864)(153.4091, -69.3864)
\put(153.4091, -124.6136){\line(1, 1){ 27.6136}}
\qbezier(157.7045, -116.6364)(163.2273, -114.7955)(163.2273, -114.7955)
\qbezier(161.3864, -120.3182)(163.2273, -114.7955)(163.2273, -114.7955)
\put( 95.1136, -124.6136){\vector(1, 0){116.5909}}
\put(125.7955, -155.2955){\vector(0, 1){168.7500}}
\put(220.9091, -129.2159){\makebox(0,0)[]{\footnotesize{$\chi$}}}
\put(116.5909,  14.9886){\makebox(0,0)[]{\footnotesize{$\eta$}}}
\put(116.8977, -14.1591){\makebox(0,0)[]{\footnotesize{$2 \pi$}}}
\put(118.1250, -69.3864){\makebox(0,0)[]{\footnotesize{$\pi$}}}
\put(181.9432, -133.8182){\makebox(0,0)[]{\footnotesize{$\pi$}}}
\put(118.1250, -40.2386){\makebox(0,0)[]{\footnotesize{$P$}}}
\put(153.4091, -132.2841){\makebox(0,0)[]{\footnotesize{$A$}}}
\put(144.2045, -106.2045){\makebox(0,0)[]{\small{$I$}}}
\put(144.2045, -32.5682){\makebox(0,0)[]{\small{$I$}}}
\put(165.6818, -61.7159){\makebox(0,0)[]{\small{$II$}}}
\end{picture}
\vspace{0mm} \newline
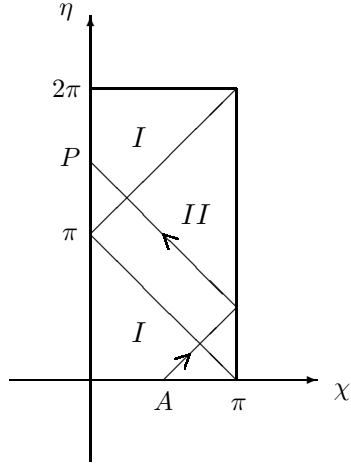
{\footnotesize \sf Figure 4. Past light cone of an event $P$ in the
$(\eta,\chi)$-plane in a closed universe model.}
\vspace{0mm} \newline
\vspace*{3mm} \newline
\begin{picture}(50,182)(64,-202)
\put(151.8750, -115.9844){\line(1, 1){ 46.4062}}
\qbezier(175.0781, -88.5625)(181.4062, -86.4531)(181.4062, -86.4531)
\qbezier(179.2969, -92.7813)(181.4062, -86.4531)(181.4062, -86.4531)
\put( 69.6094, -115.9844){\vector(1, 0){118.1250}}
\put(128.6719, -183.4844){\vector(0, 1){135.0000}}
\put(198.2812, -119.1484){\makebox(0,0)[]{\footnotesize{$R$}}}
\put(122.3437, -41.1016){\makebox(0,0)[]{\footnotesize{$T$}}}
\put(151.8750, -124.4219){\makebox(0,0)[]{\footnotesize{$A$}}}
\end{picture}
\begin{picture}(50,182)(-36,-202)
\put( 82.2656, -92.7813){\line(1, -1){ 46.4063}}
\qbezier( 99.1406, -105.4375)( 92.8125, -103.3281)( 92.8125, -103.3281)
\qbezier( 94.9219, -109.6563)( 92.8125, -103.3281)( 92.8125, -103.3281)
\put( 82.2656, -185.5937){\line(1, 1){ 46.4063}}
\qbezier( 90.7031, -172.9375)( 97.0312, -170.8281)( 97.0312, -170.8281)
\qbezier( 94.9219, -177.1562)( 97.0312, -170.8281)( 97.0312, -170.8281)
\put( 69.6094, -115.9844){\vector(1, 0){118.1250}}
\put(128.6719, -183.4844){\vector(0, 1){135.0000}}
\put(198.2812, -119.1484){\makebox(0,0)[]{\footnotesize{$\widetilde{R}$}}}
\put(122.3437, -41.1016){\makebox(0,0)[]{\footnotesize{$\widetilde{T}$}}}
\end{picture}
\begin{picture}(50,182)(-136,-202)
\put(128.6719, -139.1875){\line(1, -1){ 46.4063}}
\qbezier(151.8750, -158.1719)(145.5469, -156.0625)(145.5469, -156.0625)
\qbezier(147.6562, -162.3906)(145.5469, -156.0625)(145.5469, -156.0625)
\put( 69.6094, -115.9844){\vector(1, 0){118.1250}}
\put(128.6719, -183.4844){\vector(0, 1){135.0000}}
\put(198.2812, -119.1484){\makebox(0,0)[]{\footnotesize{$R$}}}
\put(122.3437, -41.1016){\makebox(0,0)[]{\footnotesize{$T$}}}
\put(120.2344, -139.1875){\makebox(0,0)[]{\footnotesize{$P$}}}
\end{picture}
\vspace{0mm} \newline
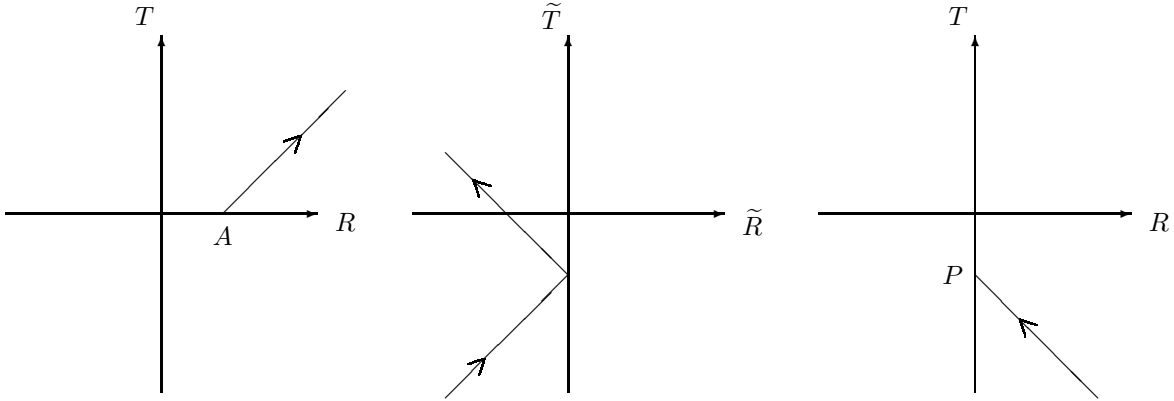
{\footnotesize \sf Figure 5. Past light cone of an event $P$ in the
conformal coordinate systems in a closed universe model.}
\vspace{0mm} \newline
\itm Again $\eta = 0$ for $t = 0$, and there is no continual creation.
On the other hand there is continual annihilation of conformal space,
as was explained by considering a similar situation in section 6 of
paper [2]. In the region ${\chi}_3 < \chi < \pi$ where
${\chi}_3 = \pi - 2 \arctan (\beta / \alpha)$,
a spherical hole where space is not defined develops in the
$(\widetilde{T},\widetilde{R})$-system as shown in Figure 6(a).
It first appears at $\widetilde{R} = 0$ at the point of time
$\widetilde{T} = \alpha / \beta$.
However, a reference particle with $0 < \chi < {\chi}_3$ escapes
the hole in the $(\widetilde{T},\widetilde{R})$-system and enters the
$(T,R)$-system as shown in Figure 6(b). All world lines end on the
hyperbola given by equation \eqref{e_11} with ${\eta}_0 = {\eta}_3$,
representing the Big Crunch. Hence a person at $R = 0$ will observe
that the Big Crunch approaches and reaches his position at
$T = - \beta / \alpha$.
\vspace{0mm} \newline
%
\begin{picture}(50,212)(14,-207)
\qbezier( 84.6773, -119.4365)( 87.2703, -116.3146)( 89.4537, -113.5235)
\qbezier( 89.4537, -113.5235)( 91.6371, -110.7324)( 93.4531, -108.2184)
\qbezier( 93.4531, -108.2184)( 95.2691, -105.7043)( 96.7526, -103.4188)
\qbezier( 96.7526, -103.4188)( 98.2361, -101.1333)( 99.4159, -99.0322)
\qbezier( 99.4159, -99.0322)(100.5956, -96.9312)(101.4942, -94.9741)
\qbezier(101.4942, -94.9741)(102.3928, -93.0170)(103.0277, -91.1661)
\qbezier(103.0277, -91.1661)(103.6626, -89.3152)(104.0460, -87.5349)
\qbezier(104.0460, -87.5349)(104.4293, -85.7545)(104.5686, -84.0104)
\qbezier(104.5686, -84.0104)(104.7078, -82.2662)(104.6056, -80.5247)
\qbezier(104.6056, -80.5247)(104.5034, -78.7831)(104.1579, -77.0105)
\qbezier(104.1579, -77.0105)(103.8123, -75.2379)(103.2166, -73.4001)
\qbezier(103.2166, -73.4001)(102.6209, -71.5623)(101.7637, -69.6238)
\qbezier(101.7637, -69.6238)(100.9065, -67.6854)( 99.7712, -65.6090)
\qbezier( 99.7712, -65.6090)( 98.6359, -63.5325)( 97.2006, -61.2781)
\qbezier( 97.2006, -61.2781)( 95.7654, -59.0236)( 94.0024, -56.5476)
\qbezier( 94.0024, -56.5476)( 92.2395, -54.0716)( 90.1149, -51.3263)
\qbezier( 90.1149, -51.3263)( 87.9904, -48.5810)( 85.4632, -45.5136)
\qbezier( 85.4632, -45.5136)( 82.9360, -42.4461)( 79.9575, -38.9973)
\qbezier( 79.9575, -38.9973)( 76.9790, -35.5485)( 73.4917, -31.6519)
\qbezier( 94.8879, -121.7489)( 96.7866, -119.3551)( 98.4649, -117.1322)
\qbezier( 98.4649, -117.1322)(100.1431, -114.9092)(101.6161, -112.8368)
\qbezier(101.6161, -112.8368)(103.0892, -110.7645)(104.3705, -108.8236)
\qbezier(104.3705, -108.8236)(105.6518, -106.8828)(106.7532, -105.0558)
\qbezier(106.7532, -105.0558)(107.8545, -103.2288)(108.7859, -101.4989)
\qbezier(108.7859, -101.4989)(109.7174, -99.7690)(110.4874, -98.1204)
\qbezier(110.4874, -98.1204)(111.2575, -96.4718)(111.8732, -94.8894)
\qbezier(111.8732, -94.8894)(112.4889, -93.3069)(112.9559, -91.7762)
\qbezier(112.9559, -91.7762)(113.4229, -90.2455)(113.7455, -88.7525)
\qbezier(113.7455, -88.7525)(114.0680, -87.2595)(114.2491, -85.7906)
\qbezier(114.2491, -85.7906)(114.4302, -84.3217)(114.4714, -82.8634)
\qbezier(114.4714, -82.8634)(114.5126, -81.4051)(114.4144, -79.9440)
\qbezier(114.4144, -79.9440)(114.3162, -78.4830)(114.0777, -77.0059)
\qbezier(114.0777, -77.0059)(113.8391, -75.5287)(113.4580, -74.0220)
\qbezier(113.4580, -74.0220)(113.0769, -72.5153)(112.5498, -70.9652)
\qbezier(112.5498, -70.9652)(112.0227, -69.4151)(111.3447, -67.8074)
\qbezier(111.3447, -67.8074)(110.6667, -66.1998)(109.8317, -64.5198)
\qbezier(109.8317, -64.5198)(108.9967, -62.8399)(107.9970, -61.0723)
\qbezier(107.9970, -61.0723)(106.9973, -59.3047)(105.8238, -57.4333)
\qbezier(107.2154, -123.3945)(108.3564, -121.5260)(109.3961, -119.7176)
\qbezier(109.3961, -119.7176)(110.4358, -117.9092)(111.3774, -116.1554)
\qbezier(111.3774, -116.1554)(112.3189, -114.4016)(113.1650, -112.6971)
\qbezier(113.1650, -112.6971)(114.0112, -110.9926)(114.7646, -109.3322)
\qbezier(114.7646, -109.3322)(115.5181, -107.6718)(116.1810, -106.0504)
\qbezier(116.1810, -106.0504)(116.8440, -104.4290)(117.4185, -102.8418)
\qbezier(117.4185, -102.8418)(117.9930, -101.2545)(118.4808, -99.6965)
\qbezier(118.4808, -99.6965)(118.9686, -98.1386)(119.3712, -96.6051)
\qbezier(119.3712, -96.6051)(119.7738, -95.0717)(120.0924, -93.5581)
\qbezier(120.0924, -93.5581)(120.4110, -92.0446)(120.6466, -90.5463)
\qbezier(120.6466, -90.5463)(120.8822, -89.0480)(121.0355, -87.5605)
\qbezier(121.0355, -87.5605)(121.1888, -86.0729)(121.2602, -84.5916)
\qbezier(121.2602, -84.5916)(121.3316, -83.1102)(121.3215, -81.6305)
\qbezier(121.3215, -81.6305)(121.3113, -80.1509)(121.2194, -78.6684)
\qbezier(121.2194, -78.6684)(121.1276, -77.1859)(120.9538, -75.6961)
\qbezier(120.9538, -75.6961)(120.7801, -74.2063)(120.5238, -72.7046)
\qbezier(120.5238, -72.7046)(120.2676, -71.2030)(119.9281, -69.6849)
\qbezier(119.9281, -69.6849)(119.5886, -68.1668)(119.1649, -66.6277)
\qbezier(119.1649, -66.6277)(118.7412, -65.0886)(118.2318, -63.5237)
\qbezier( 62.8971, -113.0551)( 64.4730, -110.7917)( 65.8267, -108.6770)
\qbezier( 65.8267, -108.6770)( 67.1805, -106.5622)( 68.3260, -104.5743)
\qbezier( 68.3260, -104.5743)( 69.4716, -102.5864)( 70.4208, -100.7048)
\qbezier( 70.4208, -100.7048)( 71.3700, -98.8232)( 72.1326, -97.0284)
\qbezier( 72.1326, -97.0284)( 72.8952, -95.2337)( 73.4792, -93.5072)
\qbezier( 73.4792, -93.5072)( 74.0631, -91.7808)( 74.4744, -90.1049)
\qbezier( 74.4744, -90.1049)( 74.8857, -88.4290)( 75.1286, -86.7863)
\qbezier( 75.1286, -86.7863)( 75.3715, -85.1436)( 75.4485, -83.5171)
\qbezier( 75.4485, -83.5171)( 75.5254, -81.8906)( 75.4373, -80.2636)
\qbezier( 75.4373, -80.2636)( 75.3492, -78.6365)( 75.0951, -76.9921)
\qbezier( 75.0951, -76.9921)( 74.8409, -75.3477)( 74.4182, -73.6689)
\qbezier( 74.4182, -73.6689)( 73.9954, -71.9902)( 73.3996, -70.2597)
\qbezier( 73.3996, -70.2597)( 72.8039, -68.5293)( 72.0289, -66.7292)
\qbezier( 72.0289, -66.7292)( 71.2540, -64.9292)( 70.2919, -63.0411)
\qbezier( 70.2919, -63.0411)( 69.3298, -61.1529)( 68.1707, -59.1571)
\qbezier( 68.1707, -59.1571)( 67.0115, -57.1613)( 65.6432, -55.0373)
\qbezier( 65.6432, -55.0373)( 64.2750, -52.9132)( 62.6835, -50.6390)
\qbezier( 62.6835, -50.6390)( 61.0921, -48.3647)( 59.2610, -45.9168)
\qbezier( 59.2610, -45.9168)( 57.4300, -43.4689)( 55.3403, -40.8220)
\thicklines
\qbezier( 73.9767, -116.1947)( 76.1246, -113.4609)( 77.9446, -110.9763)
\qbezier( 77.9446, -110.9763)( 79.7647, -108.4918)( 81.2858, -106.2171)
\qbezier( 81.2858, -106.2171)( 82.8069, -103.9424)( 84.0532, -101.8416)
\qbezier( 84.0532, -101.8416)( 85.2995, -99.7407)( 86.2907, -97.7804)
\qbezier( 86.2907, -97.7804)( 87.2819, -95.8201)( 88.0337, -93.9692)
\qbezier( 88.0337, -93.9692)( 88.7855, -92.1184)( 89.3099, -90.3477)
\qbezier( 89.3099, -90.3477)( 89.8342, -88.5770)( 90.1394, -86.8585)
\qbezier( 90.1394, -86.8585)( 90.4445, -85.1399)( 90.5354, -83.4462)
\qbezier( 90.5354, -83.4462)( 90.6262, -81.7525)( 90.5041, -80.0568)
\qbezier( 90.5041, -80.0568)( 90.3821, -78.3612)( 90.0452, -76.6367)
\qbezier( 90.0452, -76.6367)( 89.7083, -74.9122)( 89.1513, -73.1315)
\qbezier( 89.1513, -73.1315)( 88.5942, -71.3509)( 87.8082, -69.4859)
\qbezier( 87.8082, -69.4859)( 87.0222, -67.6209)( 85.9947, -65.6420)
\qbezier( 85.9947, -65.6420)( 84.9672, -63.6631)( 83.6820, -61.5389)
\qbezier( 83.6820, -61.5389)( 82.3968, -59.4147)( 80.8335, -57.1117)
\qbezier( 80.8335, -57.1117)( 79.2703, -54.8086)( 77.4041, -52.2901)
\qbezier( 77.4041, -52.2901)( 75.5380, -49.7716)( 73.3394, -46.9978)
\qbezier( 73.3394, -46.9978)( 71.1409, -44.2239)( 68.5750, -41.1509)
\qbezier( 68.5750, -41.1509)( 66.0092, -38.0778)( 63.0355, -34.6568)
\thinlines
\qbezier( 70.1674, -28.6594)( 72.6664, -30.9214)( 74.9944, -32.9929)
\qbezier( 74.9944, -32.9929)( 77.3224, -35.0644)( 79.4940, -36.9584)
\qbezier( 79.4940, -36.9584)( 81.6656, -38.8524)( 83.6942, -40.5807)
\qbezier( 83.6942, -40.5807)( 85.7229, -42.3090)( 87.6213, -43.8823)
\qbezier( 87.6213, -43.8823)( 89.5197, -45.4556)( 91.2996, -46.8839)
\qbezier( 91.2996, -46.8839)( 93.0796, -48.3121)( 94.7523, -49.6041)
\qbezier( 94.7523, -49.6041)( 96.4249, -50.8961)( 98.0007, -52.0599)
\qbezier( 98.0007, -52.0599)( 99.5764, -53.2238)(101.0651, -54.2667)
\qbezier(101.0651, -54.2667)(102.5538, -55.3096)(103.9648, -56.2382)
\qbezier(103.9648, -56.2382)(105.3757, -57.1667)(106.7176, -57.9867)
\qbezier(106.7176, -57.9867)(108.0596, -58.8066)(109.3409, -59.5230)
\qbezier(109.3409, -59.5230)(110.6222, -60.2395)(111.8509, -60.8569)
\qbezier(111.8509, -60.8569)(113.0796, -61.4743)(114.2633, -61.9965)
\qbezier(114.2633, -61.9965)(115.4471, -62.5187)(116.5932, -62.9490)
\qbezier(116.5932, -62.9490)(117.7394, -63.3794)(118.8551, -63.7204)
\qbezier(118.8551, -63.7204)(119.9708, -64.0615)(121.0630, -64.3154)
\qbezier(121.0630, -64.3154)(122.1553, -64.5694)(123.2309, -64.7378)
\qbezier(123.2309, -64.7378)(124.3064, -64.9063)(125.3721, -64.9902)
\qbezier(125.3721, -64.9902)(126.4377, -65.0742)(127.5000, -65.0742)
\put( 52.5000, -82.0000){\vector(1, 0){150.0000}}
\put(127.5000, -187.0000){\vector(0, 1){175.0000}}
\put(210.0000, -85.7500){\makebox(0,0)[]{\footnotesize{$\widetilde{R}$}}}
\put(120.0000,  -5.7500){\makebox(0,0)[]{\footnotesize{$\widetilde{T}$}}}
\put( 96.2500, -134.5000){\makebox(0,0)[]{\footnotesize{$\chi > {\chi}_3$}}}
\put( 48.7500, -123.2500){\makebox(0,0)[]{\footnotesize{$\chi < {\chi}_3$}}}
\put( 73.5000, -22.0000){\makebox(0,0)[]{\footnotesize{$\eta = {\eta}_3$}}}
\end{picture}
\begin{picture}(50,212)(-136,-207)
\qbezier(151.6485, -128.1684)(152.2237, -129.0299)(152.8206, -129.9060)
\qbezier(152.8206, -129.9060)(153.4175, -130.7822)(154.0364, -131.6735)
\qbezier(154.0364, -131.6735)(154.6553, -132.5648)(155.2966, -133.4717)
\qbezier(155.2966, -133.4717)(155.9379, -134.3787)(156.6020, -135.3020)
\qbezier(156.6020, -135.3020)(157.2661, -136.2252)(157.9534, -137.1653)
\qbezier(157.9534, -137.1653)(158.6407, -138.1054)(159.3516, -139.0629)
\qbezier(159.3516, -139.0629)(160.0626, -140.0204)(160.7976, -140.9959)
\qbezier(160.7976, -140.9959)(161.5327, -141.9715)(162.2923, -142.9656)
\qbezier(162.2923, -142.9656)(163.0518, -143.9598)(163.8364, -144.9731)
\qbezier(163.8364, -144.9731)(164.6210, -145.9865)(165.4311, -147.0198)
\qbezier(165.4311, -147.0198)(166.2412, -148.0530)(167.0773, -149.1068)
\qbezier(167.0773, -149.1068)(167.9134, -150.1606)(168.7760, -151.2355)
\qbezier(168.7760, -151.2355)(169.6387, -152.3104)(170.5283, -153.4072)
\qbezier(170.5283, -153.4072)(171.4180, -154.5039)(172.3353, -155.6232)
\qbezier(172.3353, -155.6232)(173.2526, -156.7425)(174.1980, -157.8850)
\qbezier(174.1980, -157.8850)(175.1435, -159.0275)(176.1177, -160.1939)
\qbezier(176.1177, -160.1939)(177.0920, -161.3603)(178.0956, -162.5513)
\qbezier(178.0956, -162.5513)(179.0992, -163.7423)(180.1328, -164.9588)
\qbezier(180.1328, -164.9588)(181.1664, -166.1752)(182.2306, -167.4178)
\qbezier(142.2425, -122.6976)(142.7637, -123.7829)(143.3129, -124.8819)
\qbezier(143.3129, -124.8819)(143.8621, -125.9809)(144.4396, -127.0943)
\qbezier(144.4396, -127.0943)(145.0171, -128.2077)(145.6234, -129.3363)
\qbezier(145.6234, -129.3363)(146.2296, -130.4649)(146.8650, -131.6094)
\qbezier(146.8650, -131.6094)(147.5003, -132.7539)(148.1652, -133.9151)
\qbezier(148.1652, -133.9151)(148.8300, -135.0762)(149.5249, -136.2548)
\qbezier(149.5249, -136.2548)(150.2197, -137.4333)(150.9450, -138.6301)
\qbezier(150.9450, -138.6301)(151.6702, -139.8268)(152.4264, -141.0425)
\qbezier(152.4264, -141.0425)(153.1826, -142.2583)(153.9701, -143.4937)
\qbezier(153.9701, -143.4937)(154.7576, -144.7292)(155.5771, -145.9853)
\qbezier(155.5771, -145.9853)(156.3966, -147.2414)(157.2485, -148.5188)
\qbezier(157.2485, -148.5188)(158.1004, -149.7963)(158.9853, -151.0960)
\qbezier(158.9853, -151.0960)(159.8702, -152.3957)(160.7887, -153.7186)
\qbezier(160.7887, -153.7186)(161.7072, -155.0414)(162.6599, -156.3882)
\qbezier(162.6599, -156.3882)(163.6126, -157.7349)(164.6001, -159.1066)
\qbezier(164.6001, -159.1066)(165.5876, -160.4782)(166.6105, -161.8756)
\qbezier(166.6105, -161.8756)(167.6335, -163.2729)(168.6926, -164.6970)
\qbezier(168.6926, -164.6970)(169.7517, -166.1210)(170.8476, -167.5726)
\qbezier(170.8476, -167.5726)(171.9435, -169.0243)(173.0770, -170.5045)
\qbezier(135.2318, -120.0139)(135.5886, -121.3066)(135.9697, -122.6062)
\qbezier(135.9697, -122.6062)(136.3508, -123.9058)(136.7564, -125.2127)
\qbezier(136.7564, -125.2127)(137.1620, -126.5197)(137.5921, -127.8345)
\qbezier(137.5921, -127.8345)(138.0223, -129.1493)(138.4772, -130.4725)
\qbezier(138.4772, -130.4725)(138.9322, -131.7956)(139.4120, -133.1275)
\qbezier(139.4120, -133.1275)(139.8919, -134.4594)(140.3968, -135.8005)
\qbezier(140.3968, -135.8005)(140.9017, -137.1416)(141.4319, -138.4925)
\qbezier(141.4319, -138.4925)(141.9621, -139.8433)(142.5178, -141.2044)
\qbezier(142.5178, -141.2044)(143.0734, -142.5654)(143.6547, -143.9371)
\qbezier(143.6547, -143.9371)(144.2360, -145.3088)(144.8432, -146.6917)
\qbezier(144.8432, -146.6917)(145.4503, -148.0746)(146.0836, -149.4691)
\qbezier(146.0836, -149.4691)(146.7168, -150.8636)(147.3763, -152.2702)
\qbezier(147.3763, -152.2702)(148.0358, -153.6769)(148.7218, -155.0961)
\qbezier(148.7218, -155.0961)(149.4079, -156.5154)(150.1207, -157.9478)
\qbezier(150.1207, -157.9478)(150.8335, -159.3802)(151.5733, -160.8262)
\qbezier(151.5733, -160.8262)(152.3131, -162.2722)(153.0802, -163.7324)
\qbezier(153.0802, -163.7324)(153.8473, -165.1926)(154.6419, -166.6674)
\qbezier(154.6419, -166.6674)(155.4365, -168.1422)(156.2590, -169.6323)
\qbezier(156.2590, -169.6323)(157.0814, -171.1223)(157.9320, -172.6280)
\qbezier(191.4641, -161.4004)(188.5437, -158.7087)(185.8399, -156.2538)
\qbezier(185.8399, -156.2538)(183.1361, -153.7988)(180.6296, -151.5631)
\qbezier(180.6296, -151.5631)(178.1232, -149.3274)(175.7964, -147.2950)
\qbezier(175.7964, -147.2950)(173.4695, -145.2627)(171.3058, -143.4193)
\qbezier(171.3058, -143.4193)(169.1420, -141.5760)(167.1260, -139.9085)
\qbezier(167.1260, -139.9085)(165.1100, -138.2410)(163.2275, -136.7376)
\qbezier(163.2275, -136.7376)(161.3449, -135.2342)(159.5824, -133.8841)
\qbezier(159.5824, -133.8841)(157.8199, -132.5341)(156.1649, -131.3278)
\qbezier(156.1649, -131.3278)(154.5100, -130.1216)(152.9509, -129.0506)
\qbezier(152.9509, -129.0506)(151.3918, -127.9796)(149.9175, -127.0363)
\qbezier(149.9175, -127.0363)(148.4431, -126.0930)(147.0431, -125.2706)
\qbezier(147.0431, -125.2706)(145.6430, -124.4482)(144.3074, -123.7409)
\qbezier(144.3074, -123.7409)(142.9717, -123.0337)(141.6909, -122.4365)
\qbezier(141.6909, -122.4365)(140.4101, -121.8394)(139.1752, -121.3481)
\qbezier(139.1752, -121.3481)(137.9402, -120.8568)(136.7422, -120.4679)
\qbezier(136.7422, -120.4679)(135.5443, -120.0789)(134.3749, -119.7896)
\qbezier(134.3749, -119.7896)(133.2055, -119.5003)(132.0563, -119.3086)
\qbezier(132.0563, -119.3086)(130.9072, -119.1169)(129.7701, -119.0214)
\qbezier(129.7701, -119.0214)(128.6330, -118.9258)(127.5000, -118.9258)
\put( 52.5000, -82.0000){\vector(1, 0){150.0000}}
\put(127.5000, -187.0000){\vector(0, 1){175.0000}}
\put(210.0000, -85.7500){\makebox(0,0)[]{\footnotesize{$R$}}}
\put(120.0000,  -5.7500){\makebox(0,0)[]{\footnotesize{$T$}}}
\put(172.5000, -183.2500){\makebox(0,0)[]{\footnotesize{$\chi < {\chi}_3$}}}
\put(211.2500, -162.0000){\makebox(0,0)[]{\footnotesize{$\eta = {\eta}_3$}}}
\end{picture}
\vspace{0mm} \newline
%
\hspace*{36.0mm} (a) \hspace{63.5mm} (b)
\vspace*{6mm} \newline
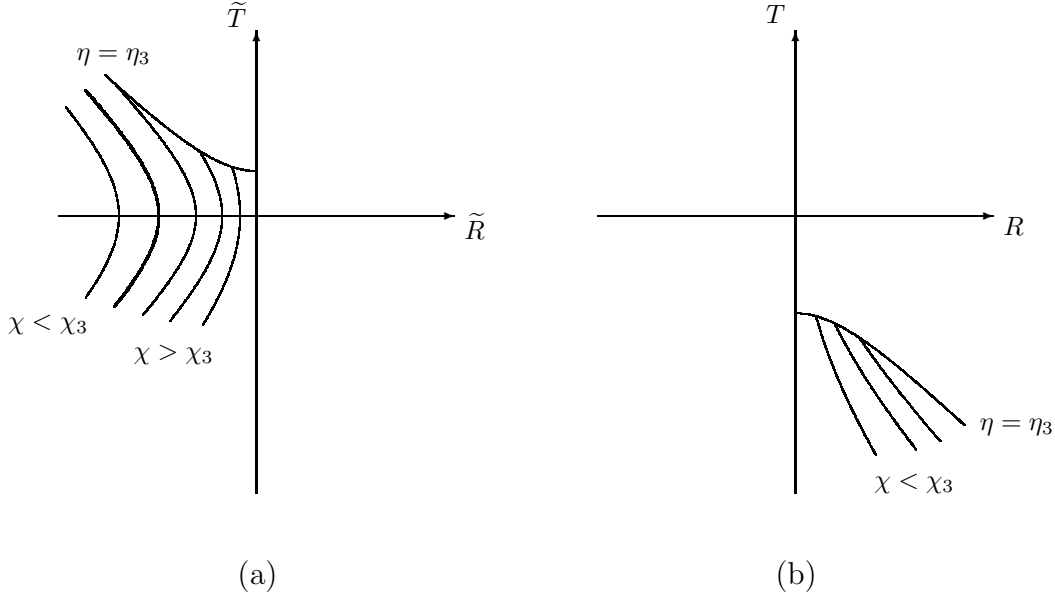
{\footnotesize \sf Figure 6. The final fate of a dust and radiation
filled universe with positive spatial curvature. (a) The curve
$\eta = {\eta}_3$ represents the boundary of an expanding spherical
hole with center at $\widetilde{R} = 0$ where conformal space disappears.
The figure shows world lines of particles with $\chi = \mbox{constant}$.
Reference particles with ${\chi}_3 < \chi < \pi$ hit the hole, and
those with $0 < \chi < {\chi}_3$ avoid it. Note that the
$\widetilde{T}$-axis corresponds to $\chi = \pi$, while the
$T$-axis corresponds to $\chi = 0$.
(b) As $\widetilde{T}$ approaches infinity, the
conformal clocks are reset to come from minus infinity, and
$\widetilde{R}$ changes from negative to positive values. The
conformal space then has a finite volume with decreasing size.
Again, the curve $\eta = {\eta}_3$ represents the boundary where
space is annihilated. The particles with $\chi = \mbox{constant}$
that avoided the hole in (a) come from infinitely far away and
vanish when they arrive at this boundary.}
\vspace{0mm} \newline
%
%
%
\vspace{5mm} \newline
{\bf 2.3. LIVE dominated universe with positive spatial curvature}
\vspace{3mm} \newline
In a universe model with $k = 1$ dominated by LIVE the scale factor is
\begin{equation} \label{e_139}
a(t) = \frl{1}{\widehat{H}_{\Lambda} \rule[-0mm]{0mm}{4.25mm}}
\cosh (\widehat{H}_{\Lambda} t)
\mbox{ ,}
\end{equation}
where $t$ goes from $- \infty$ to $\infty$, and $\widehat{H}_{\Lambda}$
is given in equation II:(47). This is a bouncing universe
model without a Big Bang. For this model the parametric time is
\begin{equation} \label{e_140}
\eta = \arcsin (\tanh (\widehat{H}_{\Lambda} t))
\mbox{ ,}
\end{equation}
where $-\pi / 2 < \eta < \pi / 2$.
The scale factor as expressed in parametric time is
\begin{equation} \label{e_141}
a(\eta) = \frl{1}{\widehat{H}_{\Lambda} \cos \eta \rule[-0mm]{0mm}{4.25mm}}
\mbox{ .}
\end{equation}
For a LIVE dominated universe model with positive spatial curvature we
shall consider three different types of conformal coordinates. The
coordinates of the first type are given in equation \eqref{e_49}.
The line element then takes the form [6,7]
\begin{equation} \label{e_143}
ds^2 = \frl{4}{\widehat{H}_{\Lambda}^2 \hs{0.6mm} [1 - (T^2 - R^2)]^2
\rule[-0mm]{0mm}{4.25mm}} ds_M^2
\mbox{ .}
\end{equation}
From equations I:(17) and \eqref{e_143} we find the conformal
scale factor
\begin{equation} \label{e_216}
A(T,R) = \frl{2 \hs{0.5mm} \mbox{\small sgn} (R)}
{\widehat{H}_{\Lambda} \hs{0.6mm} [1 - (T^2 - R^2)]
\rule[-0mm]{0mm}{4.25mm}}
\mbox{ .}
\end{equation}
The sign factor is due to the requirement that $A(T,R) > 0$, and the
fact that the denominator is positive for $T^2 - R^2 < 1$ and
negative for $T^2 - R^2 > 1$. From Figure 7 we see that the first
condition corresponds to $R > 0$ and the last condition to $R < 0$.
\vspace*{3mm} \newline
\begin{picture}(50,212)(39,-207)
\qbezier(199.2590, -55.3381)(195.5216, -59.5596)(192.3678, -63.2742)
\qbezier(192.3678, -63.2742)(189.2141, -66.9887)(186.5754, -70.2773)
\qbezier(186.5754, -70.2773)(183.9367, -73.5658)(181.7557, -76.4999)
\qbezier(181.7557, -76.4999)(179.5747, -79.4339)(177.8040, -82.0773)
\qbezier(177.8040, -82.0773)(176.0332, -84.7207)(174.6342, -87.1310)
\qbezier(174.6342, -87.1310)(173.2351, -89.5412)(172.1773, -91.7708)
\qbezier(172.1773, -91.7708)(171.1195, -94.0003)(170.3800, -96.0977)
\qbezier(170.3800, -96.0977)(169.6404, -98.1950)(169.2030, -100.2058)
\qbezier(169.2030, -100.2058)(168.7656, -102.2166)(168.6209, -104.1846)
\qbezier(168.6209, -104.1846)(168.4761, -106.1525)(168.6209, -108.1205)
\qbezier(168.6209, -108.1205)(168.7656, -110.0885)(169.2030, -112.0993)
\qbezier(169.2030, -112.0993)(169.6404, -114.1101)(170.3800, -116.2074)
\qbezier(170.3800, -116.2074)(171.1195, -118.3048)(172.1773, -120.5343)
\qbezier(172.1773, -120.5343)(173.2351, -122.7639)(174.6342, -125.1741)
\qbezier(174.6342, -125.1741)(176.0332, -127.5844)(177.8040, -130.2278)
\qbezier(177.8040, -130.2278)(179.5747, -132.8712)(181.7557, -135.8052)
\qbezier(181.7557, -135.8052)(183.9367, -138.7393)(186.5754, -142.0278)
\qbezier(186.5754, -142.0278)(189.2141, -145.3164)(192.3678, -149.0309)
\qbezier(192.3678, -149.0309)(195.5216, -152.7455)(199.2590, -156.9670)
\qbezier(194.2584, -46.6037)(189.0067, -52.2333)(184.6734, -57.0177)
\qbezier(184.6734, -57.0177)(180.3401, -61.8020)(176.7894, -65.8911)
\qbezier(176.7894, -65.8911)(173.2387, -69.9801)(170.3592, -73.5021)
\qbezier(170.3592, -73.5021)(167.4797, -77.0241)(165.1812, -80.0894)
\qbezier(165.1812, -80.0894)(162.8826, -83.1547)(161.0930, -85.8594)
\qbezier(161.0930, -85.8594)(159.3033, -88.5642)(157.9664, -90.9932)
\qbezier(157.9664, -90.9932)(156.6294, -93.4222)(155.7034, -95.6517)
\qbezier(155.7034, -95.6517)(154.7773, -97.8811)(154.2330, -99.9809)
\qbezier(154.2330, -99.9809)(153.6887, -102.0807)(153.5092, -104.1166)
\qbezier(153.5092, -104.1166)(153.3296, -106.1525)(153.5092, -108.1885)
\qbezier(153.5092, -108.1885)(153.6887, -110.2244)(154.2330, -112.3242)
\qbezier(154.2330, -112.3242)(154.7773, -114.4240)(155.7034, -116.6534)
\qbezier(155.7034, -116.6534)(156.6294, -118.8828)(157.9664, -121.3119)
\qbezier(157.9664, -121.3119)(159.3033, -123.7409)(161.0930, -126.4457)
\qbezier(161.0930, -126.4457)(162.8826, -129.1504)(165.1812, -132.2157)
\qbezier(165.1812, -132.2157)(167.4797, -135.2810)(170.3592, -138.8030)
\qbezier(170.3592, -138.8030)(173.2387, -142.3250)(176.7894, -146.4140)
\qbezier(176.7894, -146.4140)(180.3401, -150.5031)(184.6734, -155.2874)
\qbezier(184.6734, -155.2874)(189.0067, -160.0718)(194.2584, -165.7014)
\qbezier(160.8232, -66.2522)(158.4018, -69.2079)(156.3225, -71.8905)
\qbezier(156.3225, -71.8905)(154.2433, -74.5732)(152.4754, -77.0223)
\qbezier(152.4754, -77.0223)(150.7076, -79.4714)(149.2250, -81.7233)
\qbezier(149.2250, -81.7233)(147.7425, -83.9752)(146.5233, -86.0632)
\qbezier(146.5233, -86.0632)(145.3042, -88.1511)(144.3303, -90.1059)
\qbezier(144.3303, -90.1059)(143.3565, -92.0608)(142.6136, -93.9114)
\qbezier(142.6136, -93.9114)(141.8708, -95.7621)(141.3478, -97.5359)
\qbezier(141.3478, -97.5359)(140.8249, -99.3098)(140.5142, -101.0330)
\qbezier(140.5142, -101.0330)(140.2035, -102.7563)(140.1005, -104.4544)
\qbezier(140.1005, -104.4544)(139.9974, -106.1525)(140.1005, -107.8507)
\qbezier(140.1005, -107.8507)(140.2035, -109.5488)(140.5142, -111.2721)
\qbezier(140.5142, -111.2721)(140.8249, -112.9953)(141.3478, -114.7692)
\qbezier(141.3478, -114.7692)(141.8708, -116.5430)(142.6136, -118.3937)
\qbezier(142.6136, -118.3937)(143.3565, -120.2443)(144.3303, -122.1992)
\qbezier(144.3303, -122.1992)(145.3042, -124.1540)(146.5233, -126.2419)
\qbezier(146.5233, -126.2419)(147.7425, -128.3299)(149.2250, -130.5817)
\qbezier(149.2250, -130.5817)(150.7076, -132.8336)(152.4754, -135.2828)
\qbezier(152.4754, -135.2828)(154.2433, -137.7319)(156.3225, -140.4145)
\qbezier(156.3225, -140.4145)(158.4018, -143.0972)(160.8232, -146.0529)
\qbezier(138.0840, -81.2020)(137.5298, -82.5921)(137.0368, -83.9589)
\qbezier(137.0368, -83.9589)(136.5437, -85.3257)(136.1110, -86.6719)
\qbezier(136.1110, -86.6719)(135.6783, -88.0181)(135.3049, -89.3464)
\qbezier(135.3049, -89.3464)(134.9316, -90.6747)(134.6169, -91.9877)
\qbezier(134.6169, -91.9877)(134.3023, -93.3008)(134.0457, -94.6011)
\qbezier(134.0457, -94.6011)(133.7891, -95.9014)(133.5900, -97.1917)
\qbezier(133.5900, -97.1917)(133.3910, -98.4819)(133.2491, -99.7645)
\qbezier(133.2491, -99.7645)(133.1072, -101.0472)(133.0221, -102.3248)
\qbezier(133.0221, -102.3248)(132.9371, -103.6024)(132.9088, -104.8775)
\qbezier(132.9088, -104.8775)(132.8805, -106.1525)(132.9088, -107.4276)
\qbezier(132.9088, -107.4276)(132.9371, -108.7027)(133.0221, -109.9803)
\qbezier(133.0221, -109.9803)(133.1072, -111.2579)(133.2491, -112.5406)
\qbezier(133.2491, -112.5406)(133.3910, -113.8232)(133.5900, -115.1134)
\qbezier(133.5900, -115.1134)(133.7891, -116.4036)(134.0457, -117.7040)
\qbezier(134.0457, -117.7040)(134.3023, -119.0043)(134.6169, -120.3173)
\qbezier(134.6169, -120.3173)(134.9316, -121.6304)(135.3049, -122.9587)
\qbezier(135.3049, -122.9587)(135.6783, -124.2870)(136.1110, -125.6332)
\qbezier(136.1110, -125.6332)(136.5437, -126.9794)(137.0368, -128.3462)
\qbezier(137.0368, -128.3462)(137.5298, -129.7130)(138.0840, -131.1030)
\qbezier(128.1356, -83.2712)(129.1586, -83.2712)(130.1856, -83.1795)
\qbezier(128.1356, -129.0339)(129.1586, -129.0339)(130.1856, -129.1255)
\qbezier(130.1856, -83.1795)(131.2127, -83.0879)(132.2520, -82.9039)
\qbezier(130.1856, -129.1255)(131.2127, -129.2172)(132.2520, -129.4012)
\qbezier(132.2520, -82.9039)(133.2914, -82.7198)(134.3515, -82.4419)
\qbezier(132.2520, -129.4012)(133.2914, -129.5853)(134.3515, -129.8632)
\qbezier(134.3515, -82.4419)(135.4115, -82.1640)(136.5007, -81.7901)
\qbezier(134.3515, -129.8632)(135.4115, -130.1411)(136.5007, -130.5150)
\qbezier(136.5007, -81.7901)(137.5898, -81.4161)(138.7169, -80.9430)
\qbezier(136.5007, -130.5150)(137.5898, -130.8890)(138.7169, -131.3621)
\qbezier(138.7169, -80.9430)(139.8439, -80.4700)(141.0179, -79.8940)
\qbezier(138.7169, -131.3621)(139.8439, -131.8351)(141.0179, -132.4110)
\qbezier(141.0179, -79.8940)(142.1918, -79.3181)(143.4220, -78.6347)
\qbezier(141.0179, -132.4110)(142.1918, -132.9870)(143.4220, -133.6704)
\qbezier(143.4220, -78.6347)(144.6523, -77.9513)(145.9487, -77.1549)
\qbezier(143.4220, -133.6704)(144.6523, -134.3538)(145.9487, -135.1502)
\qbezier(145.9487, -77.1549)(147.2451, -76.3585)(148.6180, -75.4428)
\qbezier(145.9487, -135.1502)(147.2451, -135.9465)(148.6180, -136.8622)
\qbezier(148.6180, -75.4428)(149.9909, -74.5271)(151.4514, -73.4848)
\qbezier(148.6180, -136.8622)(149.9909, -137.7780)(151.4514, -138.8203)
\qbezier(151.4514, -73.4848)(152.9119, -72.4424)(154.4716, -71.2650)
\qbezier(151.4514, -138.8203)(152.9119, -139.8627)(154.4716, -141.0401)
\qbezier(154.4716, -71.2650)(156.0313, -70.0876)(157.7028, -68.7657)
\qbezier(154.4716, -141.0401)(156.0313, -142.2175)(157.7028, -143.5394)
\qbezier(157.7028, -68.7657)(159.3743, -67.4438)(161.1708, -65.9669)
\qbezier(157.7028, -143.5394)(159.3743, -144.8613)(161.1708, -146.3382)
\qbezier(161.1708, -65.9669)(162.9674, -64.4900)(164.9035, -62.8462)
\qbezier(161.1708, -146.3382)(162.9674, -147.8151)(164.9035, -149.4588)
\qbezier(164.9035, -62.8462)(166.8396, -61.2025)(168.9307, -59.3786)
\qbezier(164.9035, -149.4588)(166.8396, -151.1026)(168.9307, -152.9265)
\qbezier(168.9307, -59.3786)(171.0219, -57.5548)(173.2848, -55.5363)
\qbezier(168.9307, -152.9265)(171.0219, -154.7503)(173.2848, -156.7688)
\qbezier(173.2848, -55.5363)(175.5477, -53.5178)(178.0005, -51.2885)
\qbezier(173.2848, -156.7688)(175.5477, -158.7873)(178.0005, -161.0166)
\qbezier(178.0005, -51.2885)(180.4533, -49.0592)(183.1156, -46.6012)
\qbezier(178.0005, -161.0166)(180.4533, -163.2459)(183.1156, -165.7039)
\qbezier(183.1156, -46.6012)(185.7780, -44.1432)(188.6712, -41.4368)
\qbezier(183.1156, -165.7039)(185.7780, -168.1619)(188.6712, -170.8682)
\put( 48.0508, -106.1525){\vector(1, 0){160.1695}}
\put(128.1356, -186.2373){\vector(0, 1){160.1695}}
\put(215.0847, -109.5847){\makebox(0,0)[]{\footnotesize{$R$}}}
\put(121.2712, -24.9237){\makebox(0,0)[]{\footnotesize{$T$}}}
\put(227.6695, -156.4915){\makebox(0,0)[]{\footnotesize{$\chi = \mbox{const}$}}}
\put(207.0763, -35.2203){\makebox(0,0)[]{\small{$\eta = \frac{\pi}{2}$}}}
\put(213.9407, -177.0847){\makebox(0,0)[]{\small{$\eta = - \frac{\pi}{2}$}}}
\end{picture}
\begin{picture}(50,212)(-171,-207)
\qbezier(121.2018, -82.2437)(120.9304, -81.3077)(120.6377, -80.3654)
\qbezier(121.2018, -130.0614)(120.9304, -130.9973)(120.6377, -131.9397)
\qbezier(120.6377, -80.3654)(120.3449, -79.4230)(120.0307, -78.4738)
\qbezier(120.6377, -131.9397)(120.3449, -132.8821)(120.0307, -133.8313)
\qbezier(120.0307, -78.4738)(119.7164, -77.5245)(119.3806, -76.5679)
\qbezier(120.0307, -133.8313)(119.7164, -134.7806)(119.3806, -135.7372)
\qbezier(119.3806, -76.5679)(119.0447, -75.6113)(118.6870, -74.6468)
\qbezier(119.3806, -135.7372)(119.0447, -136.6938)(118.6870, -137.6583)
\qbezier(118.6870, -74.6468)(118.3293, -73.6823)(117.9497, -72.7094)
\qbezier(118.6870, -137.6583)(118.3293, -138.6228)(117.9497, -139.5957)
\qbezier(117.9497, -72.7094)(117.5700, -71.7365)(117.1681, -70.7548)
\qbezier(117.9497, -139.5957)(117.5700, -140.5686)(117.1681, -141.5503)
\qbezier(117.1681, -70.7548)(116.7663, -69.7730)(116.3420, -68.7819)
\qbezier(117.1681, -141.5503)(116.7663, -142.5321)(116.3420, -143.5232)
\qbezier(116.3420, -68.7819)(115.9177, -67.7908)(115.4709, -66.7898)
\qbezier(116.3420, -143.5232)(115.9177, -144.5143)(115.4709, -145.5153)
\qbezier(115.4709, -66.7898)(115.0240, -65.7888)(114.5543, -64.7773)
\qbezier(115.4709, -145.5153)(115.0240, -146.5163)(114.5543, -147.5278)
\qbezier(114.5543, -64.7773)(114.0845, -63.7659)(113.5917, -62.7436)
\qbezier(114.5543, -147.5278)(114.0845, -148.5392)(113.5917, -149.5615)
\qbezier(113.5917, -62.7436)(113.0989, -61.7212)(112.5827, -60.6874)
\qbezier(113.5917, -149.5615)(113.0989, -150.5839)(112.5827, -151.6177)
\qbezier(112.5827, -60.6874)(112.0666, -59.6536)(111.5268, -58.6078)
\qbezier(112.5827, -151.6177)(112.0666, -152.6515)(111.5268, -153.6973)
\qbezier(111.5268, -58.6078)(110.9871, -57.5620)(110.4234, -56.5037)
\qbezier(111.5268, -153.6973)(110.9871, -154.7431)(110.4234, -155.8014)
\qbezier(110.4234, -56.5037)(109.8597, -55.4454)(109.2719, -54.3740)
\qbezier(110.4234, -155.8014)(109.8597, -156.8597)(109.2719, -157.9311)
\qbezier(109.2719, -54.3740)(108.6841, -53.3026)(108.0718, -52.2176)
\qbezier(109.2719, -157.9311)(108.6841, -159.0025)(108.0718, -160.0875)
\qbezier(108.0718, -52.2176)(107.4595, -51.1326)(106.8223, -50.0334)
\qbezier(108.0718, -160.0875)(107.4595, -161.1725)(106.8223, -162.2717)
\qbezier(106.8223, -50.0334)(106.1852, -48.9342)(105.5230, -47.8202)
\qbezier(106.8223, -162.2717)(106.1852, -163.3709)(105.5230, -164.4848)
\qbezier(105.5230, -47.8202)(104.8608, -46.7063)(104.1731, -45.5770)
\qbezier(105.5230, -164.4848)(104.8608, -165.5988)(104.1731, -166.7281)
\qbezier(104.1731, -45.5770)(103.4854, -44.4477)(102.7718, -43.3025)
\qbezier(104.1731, -166.7281)(103.4854, -167.8574)(102.7718, -169.0025)
\qbezier(110.5346, -77.2847)(110.0776, -76.5352)(109.5967, -75.7710)
\qbezier(110.5346, -135.0204)(110.0776, -135.7698)(109.5967, -136.5341)
\qbezier(109.5967, -75.7710)(109.1158, -75.0067)(108.6105, -74.2268)
\qbezier(109.5967, -136.5341)(109.1158, -137.2984)(108.6105, -138.0783)
\qbezier(108.6105, -74.2268)(108.1051, -73.4469)(107.5748, -72.6506)
\qbezier(108.6105, -138.0783)(108.1051, -138.8582)(107.5748, -139.6545)
\qbezier(107.5748, -72.6506)(107.0445, -71.8543)(106.4888, -71.0409)
\qbezier(107.5748, -139.6545)(107.0445, -140.4508)(106.4888, -141.2642)
\qbezier(106.4888, -71.0409)(105.9330, -70.2274)(105.3512, -69.3959)
\qbezier(106.4888, -141.2642)(105.9330, -142.0777)(105.3512, -142.9092)
\qbezier(105.3512, -69.3959)(104.7694, -68.5644)(104.1610, -67.7141)
\qbezier(105.3512, -142.9092)(104.7694, -143.7407)(104.1610, -144.5910)
\qbezier(104.1610, -67.7141)(103.5525, -66.8638)(102.9169, -65.9938)
\qbezier(104.1610, -144.5910)(103.5525, -145.4413)(102.9169, -146.3113)
\qbezier(102.9169, -65.9938)(102.2812, -65.1237)(101.6177, -64.2332)
\qbezier(102.9169, -146.3113)(102.2812, -147.1813)(101.6177, -148.0719)
\qbezier(101.6177, -64.2332)(100.9542, -63.3426)(100.2621, -62.4306)
\qbezier(101.6177, -148.0719)(100.9542, -148.9625)(100.2621, -149.8745)
\qbezier(100.2621, -62.4306)( 99.5700, -61.5185)( 98.8488, -60.5841)
\qbezier(100.2621, -149.8745)( 99.5700, -150.7866)( 98.8488, -151.7210)
\qbezier( 98.8488, -60.5841)( 98.1275, -59.6497)( 97.3762, -58.6920)
\qbezier( 98.8488, -151.7210)( 98.1275, -152.6554)( 97.3762, -153.6131)
\qbezier( 97.3762, -58.6920)( 96.6250, -57.7343)( 95.8431, -56.7523)
\qbezier( 97.3762, -153.6131)( 96.6250, -154.5708)( 95.8431, -155.5528)
\qbezier( 95.8431, -56.7523)( 95.0611, -55.7703)( 94.2477, -54.7631)
\qbezier( 95.8431, -155.5528)( 95.0611, -156.5348)( 94.2477, -157.5420)
\qbezier( 94.2477, -54.7631)( 93.4343, -53.7558)( 92.5885, -52.7223)
\qbezier( 94.2477, -157.5420)( 93.4343, -158.5492)( 92.5885, -159.5827)
\qbezier( 92.5885, -52.7223)( 91.7428, -51.6888)( 90.8639, -50.6281)
\qbezier( 92.5885, -159.5827)( 91.7428, -160.6162)( 90.8639, -161.6770)
\qbezier( 90.8639, -50.6281)( 89.9850, -49.5673)( 89.0721, -48.4781)
\qbezier( 90.8639, -161.6770)( 89.9850, -162.7378)( 89.0721, -163.8270)
\qbezier( 89.0721, -48.4781)( 88.1592, -47.3889)( 87.2114, -46.2703)
\qbezier( 89.0721, -163.8270)( 88.1592, -164.9161)( 87.2114, -166.0348)
\qbezier( 87.2114, -46.2703)( 86.2635, -45.1517)( 85.2798, -44.0025)
\qbezier( 87.2114, -166.0348)( 86.2635, -167.1534)( 85.2798, -168.3026)
\qbezier( 85.2798, -44.0025)( 84.2960, -42.8533)( 83.2754, -41.6724)
\qbezier( 85.2798, -168.3026)( 84.2960, -169.4517)( 83.2754, -170.6327)
\qbezier( 67.5999, -41.4368)( 70.4932, -44.1432)( 73.1556, -46.6012)
\qbezier( 67.5999, -170.8682)( 70.4932, -168.1619)( 73.1556, -165.7039)
\qbezier( 73.1556, -46.6012)( 75.8179, -49.0592)( 78.2707, -51.2885)
\qbezier( 73.1556, -165.7039)( 75.8179, -163.2459)( 78.2707, -161.0166)
\qbezier( 78.2707, -51.2885)( 80.7235, -53.5178)( 82.9864, -55.5363)
\qbezier( 78.2707, -161.0166)( 80.7235, -158.7873)( 82.9864, -156.7688)
\qbezier( 82.9864, -55.5363)( 85.2493, -57.5548)( 87.3405, -59.3786)
\qbezier( 82.9864, -156.7688)( 85.2493, -154.7503)( 87.3405, -152.9265)
\qbezier( 87.3405, -59.3786)( 89.4316, -61.2025)( 91.3677, -62.8462)
\qbezier( 87.3405, -152.9265)( 89.4316, -151.1026)( 91.3677, -149.4588)
\qbezier( 91.3677, -62.8462)( 93.3038, -64.4900)( 95.1004, -65.9669)
\qbezier( 91.3677, -149.4588)( 93.3038, -147.8151)( 95.1004, -146.3382)
\qbezier( 95.1004, -65.9669)( 96.8969, -67.4438)( 98.5684, -68.7657)
\qbezier( 95.1004, -146.3382)( 96.8969, -144.8613)( 98.5684, -143.5394)
\qbezier( 98.5684, -68.7657)(100.2398, -70.0876)(101.7996, -71.2650)
\qbezier( 98.5684, -143.5394)(100.2398, -142.2175)(101.7996, -141.0401)
\qbezier(101.7996, -71.2650)(103.3593, -72.4424)(104.8198, -73.4848)
\qbezier(101.7996, -141.0401)(103.3593, -139.8627)(104.8198, -138.8203)
\qbezier(104.8198, -73.4848)(106.2802, -74.5271)(107.6532, -75.4428)
\qbezier(104.8198, -138.8203)(106.2802, -137.7780)(107.6532, -136.8622)
\qbezier(107.6532, -75.4428)(109.0261, -76.3585)(110.3225, -77.1549)
\qbezier(107.6532, -136.8622)(109.0261, -135.9465)(110.3225, -135.1502)
\qbezier(110.3225, -77.1549)(111.6189, -77.9513)(112.8492, -78.6347)
\qbezier(110.3225, -135.1502)(111.6189, -134.3538)(112.8492, -133.6704)
\qbezier(112.8492, -78.6347)(114.0794, -79.3181)(115.2533, -79.8940)
\qbezier(112.8492, -133.6704)(114.0794, -132.9870)(115.2533, -132.4110)
\qbezier(115.2533, -79.8940)(116.4273, -80.4700)(117.5543, -80.9430)
\qbezier(115.2533, -132.4110)(116.4273, -131.8351)(117.5543, -131.3621)
\qbezier(117.5543, -80.9430)(118.6814, -81.4161)(119.7705, -81.7901)
\qbezier(117.5543, -131.3621)(118.6814, -130.8890)(119.7705, -130.5150)
\qbezier(119.7705, -81.7901)(120.8597, -82.1640)(121.9197, -82.4419)
\qbezier(119.7705, -130.5150)(120.8597, -130.1411)(121.9197, -129.8632)
\qbezier(121.9197, -82.4419)(122.9798, -82.7198)(124.0191, -82.9039)
\qbezier(121.9197, -129.8632)(122.9798, -129.5853)(124.0191, -129.4012)
\qbezier(124.0191, -82.9039)(125.0585, -83.0879)(126.0856, -83.1795)
\qbezier(124.0191, -129.4012)(125.0585, -129.2172)(126.0856, -129.1255)
\qbezier(126.0856, -83.1795)(127.1126, -83.2712)(128.1356, -83.2712)
\qbezier(126.0856, -129.1255)(127.1126, -129.0339)(128.1356, -129.0339)
\put( 48.0508, -106.1525){\vector(1, 0){160.1695}}
\put(128.1356, -186.2373){\vector(0, 1){160.1695}}
\put(215.0847, -109.5847){\makebox(0,0)[]{\footnotesize{$\widetilde{R}$}}}
\put(121.2712, -24.9237){\makebox(0,0)[]{\footnotesize{$\widetilde{T}$}}}
\put( 97.2458, -180.5169){\makebox(0,0)[]{\footnotesize{$\chi = \mbox{const}$}}}
\put( 46.4492, -35.2203){\makebox(0,0)[]{\small{$\eta = - \frac{\pi}{2}$}}}
\put( 50.1102, -177.0847){\makebox(0,0)[]{\small{$\eta = \frac{\pi}{2}$}}}
\end{picture}
\vspace{-6mm} \newline
\vspace*{-3mm} \newline
\hspace*{27.0mm} (a) \hspace{85.5mm} (b)
\vspace*{6mm} \newline
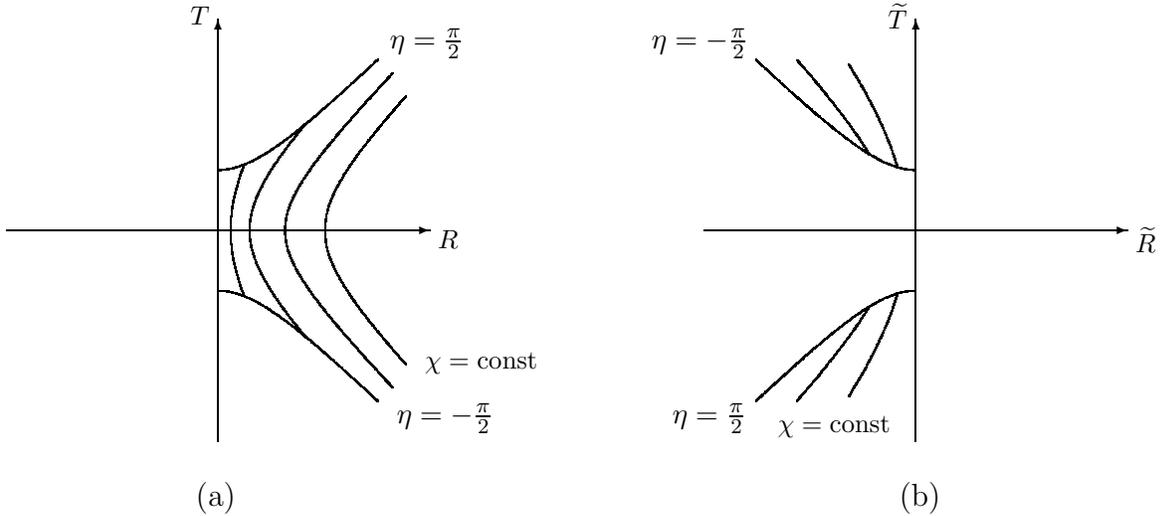
{\footnotesize \sf Figure 7. In the CFS space of a LIVE dominated universe
with $k = 1$ there is initially an empty bubble of decreasing size in which
space does not exist. New space is created, and at the conformal time
$T = -1$ the bubble vanishes. A particle with $\chi = {\chi}_1 < \pi / 2$
comes from the boundary of the bubble in region I at a point of time
$T_1 = - \sqrt{1 + R_1^2}$. It moves towards the origin, comes
instantaneously to rest at $T = 0$ and then accelerates outwards.
As shown in part (a) of the figure a new bubble is created at $T = 1$
and expands. Conformal space is annihilated at the boundary of the bubble.
The particle hits the bubble at $T = \sqrt{1 + R_1^2}$.
A particle with $\chi = {\chi}_2$ where $\pi / 2 < {\chi}_2 < \pi$ comes
from a different region II in spacetime, covered by the
$(\widetilde{T},\widetilde{R})$-coordinate system shown in part (b) of
the figure. A bubble of conformal space appears at $\widetilde{T} = 1$
and expands. The particle comes from the boundary of this bubble. It
accelerates away from the origin and moves towards infinity. Then it
enters the region I covered by the $(T,R)$-system. In this system it
comes from spatial infinity at an infinitely past time, moves to a
finite distance from the origin, and then accelerates back to spatial
infinity in the infinite future. Then it enters the region II again,
reappearing at infinite past in the $(\widetilde{T},\widetilde{R})$-system
from a direction opposite to the one it started its motion. Conformal
space is annihilated and vanishes at $\widetilde{T} = -1$. The particle
hits the boundary of space at $\widetilde{T} = - \sqrt{1 + R_2^2}$.}
\newpage
%
\makebox[8mm]{}
The parametric time $\eta$ is given in equation \eqref{e_140}
with $- \infty < t < \infty$ corresponding to
$- \pi / 2 < \eta < \pi / 2$. The coordinate region I in Figure 8
defined by $- \pi / 2 < \eta < \pi / 2$ and $0 < \chi < \pi - |\eta|$
in the $(\eta,\chi)$-system is transformed onto the the region
given by $R > 0$ and $T^2 < 1 + R^2$ in the $(T,R)$-system as shown
in Figure 7a. The coordinate region II defined by
$- \pi / 2 < \eta < \pi / 2$ and $\pi - |\eta| < \chi < \pi$
in the $(\eta,\chi)$-system is transformed onto the region given by
$\widetilde{R} < 0$ and $\widetilde{T}^2 > 1 + \widetilde{R}^2$ in
the $(\widetilde{T},\widetilde{R})$-system as shown in Figure 7b.
\vspace*{3mm} \newline
\begin{picture}(50,172)(-66,-152)
\put(158.8235, -37.4412){\line(1, -1){ 35.7353}}
\put(158.8235, -108.9118){\line(1, 1){ 35.7353}}
\put(123.0882, -37.4412){\line(1, 0){ 71.4706}}
\put(123.0882, -108.9118){\line(1, 0){ 71.4706}}
\put(194.5588, -108.9118){\line(0, 1){ 71.4706}}
\put( 83.3824, -73.1765){\vector(1, 0){150.8824}}
\put(123.0882, -132.7353){\vector(0, 1){119.1176}}
\put(242.2059, -79.1324){\makebox(0,0)[]{\footnotesize{$\chi$}}}
\put(115.1471, -11.6324){\makebox(0,0)[]{\footnotesize{$\eta$}}}
\put(113.1618, -37.4412){\makebox(0,0)[]{\small{$\frac{\pi}{2}$}}}
\put(109.1912, -108.9118){\makebox(0,0)[]{\small{$-\frac{\pi}{2}$}}}
\put(203.6912, -81.1176){\makebox(0,0)[]{\small{$\pi$}}}
\put(146.9118, -49.3529){\makebox(0,0)[]{\small{$I$}}}
\put(186.6176, -49.3529){\makebox(0,0)[]{\small{$II$}}}
\put(186.6176, -97.0000){\makebox(0,0)[]{\small{$II$}}}
\end{picture}
\vspace{0mm} \newline
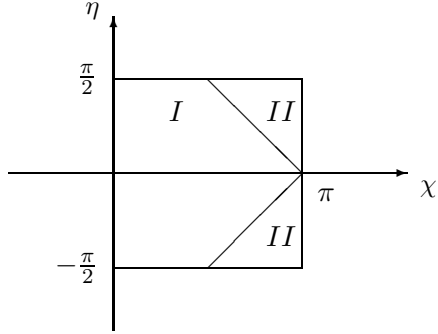
{\footnotesize \sf Figure 8. The $(\eta,\chi)$-coordinate regions of a
LIVE dominated universe with positive curvature that are transformed into
the $(T,R)$- and $(\widetilde{T},\widetilde{R})$-coordinate regions of
Figure 7.}
\vspace{0mm} \newline
\itm We shall describe the evolution of a region where
$0 < \chi < \pi / 2$ in a LIVE dominated universe with positive
spatial curvature. Initially there is a hole in a conformal space
which is arbitrarily large in the infinitely remote past.
At the boundary of the hole, the cosmic
time $t$ approaches $- \infty$. The hole decreases in size with
continual creation of new space and vacuum energy at the boundary.
This is similar to the situation described in section 5 of paper [2]. At
the conformal time $T = -1$ the hole vanishes. Then at $T = 1$ the hole
reappears, and there is continual annihilation. In this region every
reference particle with $\chi = \mbox{constant}$ is created at the
boundary of the first hole, and is annihilated at the boundary of
the second hole.
In the region with $\pi / 2 < \chi < \pi$ a conformal space
of finite extension, initially created at $\widetilde{T} = 1$ and
$\widetilde{R} = 0$, is expanding with continual creation.
%
%
%
\vspace{5mm} \newline
{\bf 2.4. New types of conformal coordinates for a LIVE dominated
universe with positive spatial curvature}
\vspace{3mm} \newline
V.F.Mukhanov [8] has introduced a parametric time
$\widehat{\eta} = \eta - \pi / 2$ where $- \pi < \widehat{\eta} < 0$.
In this way he obtains the same value $\widehat{\eta} = 0$
at the end of the universe for all cases $k = -1$, $k = 0$ and $k = 1$.
However, our choice \eqref{e_140} is natural if one focuses on the
symmetry of the function $a(t)$, i.e. on the symmetry of the expansion
history of the universe.
\itm V.F.Mukhanov uses the transformation \eqref{e_49} with $\eta$
replaced by $\widehat{\eta}$. Expressed by our coordinate $\eta$ this
corresponds to choosing $a = \pi / 2$, $b = 0$, $c = 1$ and $d = 0$
in equation I:(33). This defines a second type of conformal
coordinates for LIVE dominated universe model with positive spatial
curvature with generating function
\begin{equation} \label{e_227}
f(x) = \tan (x / 2 - \pi / 4) =
\frl{\tan (x / 2) - 1}{\tan (x / 2) + 1}
\mbox{ .}
\end{equation}
so that transformation I:(28) takes the form
\begin{equation} \label{e_228}
\widehat{T} = - \frac{\cos \eta}{\sin \eta + \cos \chi}
\mbox{\hspace{2mm} , \hspace{3mm}}
\widehat{R} = \frac{\sin \chi}{\sin \eta + \cos \chi}
\mbox{ ,}
\end{equation}
transforming the triangle defined by $0 < \chi < \pi$ and
$- \pi / 2 < \eta < \chi - \pi / 2$ onto the second quadrant
$\widehat{R} < 0$, $\widehat{T} > 0$, and the triangle defined by
$0 < \chi < \pi$ and $\chi - \pi / 2 < \eta < \pi / 2$ onto the
fourth quadrant $\widehat{R} > 0$, $\widehat{T} < 0$.
The inverse transformation is
\begin{equation} \label{e_233}
\tan \eta = \frl{(\widehat{T}^{2} - \widehat{R}^{2}) - 1}
{2 \widehat{T} \rule[-0mm]{0mm}{4.25mm}}
\mbox{\hspace{2mm} , \hspace{3mm}}
\cot \chi = \frl{(\widehat{T}^{2} - \widehat{R}^{2}) + 1}
{2 \widehat{R} \rule[-0mm]{0mm}{4.25mm}}
\mbox{ .}
\end{equation}
\itm One may wonder if it is possible to define a new type of conformal
coordinates by choosing $a = -\pi / 2$, $b = 0$, $c = 1$ and
$d = 0$ in equation I:(33). Then the generating function would be
\begin{equation} \label{e_229}
f(x) = \tan (x / 2 + \pi / 4) =
\frl{1 + \tan (x / 2)}{1 - \tan (x / 2)}
\mbox{ .}
\end{equation}
so that the transformation I:(28) would take the form
\begin{equation} \label{e_230}
\widetilde{T} = \frac{\cos \eta}{\cos \chi - \sin \eta}
\mbox{\hspace{2mm} , \hspace{3mm}}
\widetilde{R} = \frac{\sin \chi}{\cos \chi - \sin \eta}
\mbox{ ,}
\end{equation}
Replacing $\chi$ by $\pi - \chi$ and $R$ by $-R$, this transformation
has the same form as in \eqref{e_49}. Changing the sign of $R$
corresponds to replacing the coordinate $\theta$ by $\pi - \theta$
and $\phi$ by $\phi + \pi$. This is an antipodal transformation,
i.e. a reflection through the origin in $S^3$. Hence nothing new
is obtained physically.
%
%
%
The CFS coordinate $\widehat{T}$ has the value zero at the beginning
and the end of the universe. Again we see that the CFS time coordinate
behaves strangely. In the limit that the CFS time approaches infinity,
it is reset to come from minus infinity as shown in Figure 9.
\itm In order to obtain a correct physical interpretation of this figure,
one must take account of the difference between coordinate distances and
physical distances in the radial direction, $dl = A(R,T) dR$.
From equations \eqref{e_141}, I:(29) and \eqref{e_228} the scale
factor takes the form
\begin{equation} \label{e_232}
A(T,R) = - \frl{\mbox{\small sgn} R}{\widehat{H}_{\Lambda} T}
\mbox{ .}
\end{equation}
where we have omitted the hat and the tilde on $T$ and $R$.
\itm Looking at Figure 9, we have a coordinate description of particles
with constant $\chi$ in the $(\widehat{T},\widehat{R})$-system.
\itm It may be noted that the transformation equation \eqref{e_228}
can also be used for universe models with $k = 1$ containing dust and
radiation, but with different coordinate domains.
\vspace*{3mm} \newline
\begin{picture}(50,212)(-100,-207)
\qbezier(168.5484, -106.1525)(168.5484, -107.1212)(168.6188, -108.0923)
\qbezier(168.6188, -108.0923)(168.6893, -109.0635)(168.8305, -110.0424)
\qbezier(168.8305, -110.0424)(168.9717, -111.0213)(169.1844, -112.0130)
\qbezier(169.1844, -112.0130)(169.3971, -113.0047)(169.6825, -114.0145)
\qbezier(169.6825, -114.0145)(169.9679, -115.0243)(170.3274, -116.0575)
\qbezier(170.3274, -116.0575)(170.6870, -117.0907)(171.1226, -118.1527)
\qbezier(171.1226, -118.1527)(171.5582, -119.2148)(172.0722, -120.3113)
\qbezier(172.0722, -120.3113)(172.5861, -121.4079)(173.1812, -122.5447)
\qbezier(173.1812, -122.5447)(173.7762, -123.6815)(174.4554, -124.8645)
\qbezier(174.4554, -124.8645)(175.1347, -126.0476)(175.9017, -127.2831)
\qbezier(175.9017, -127.2831)(176.6687, -128.5187)(177.5276, -129.8132)
\qbezier(177.5276, -129.8132)(178.3865, -131.1078)(179.3418, -132.4682)
\qbezier(179.3418, -132.4682)(180.2970, -133.8286)(181.3537, -135.2621)
\qbezier(181.3537, -135.2621)(182.4103, -136.6955)(183.5740, -138.2096)
\qbezier(183.5740, -138.2096)(184.7377, -139.7236)(186.0145, -141.3263)
\qbezier(186.0145, -141.3263)(187.2913, -142.9289)(188.6880, -144.6286)
\qbezier(188.6880, -144.6286)(190.0846, -146.3283)(191.6086, -148.1340)
\qbezier(191.6086, -148.1340)(193.1325, -149.9397)(194.7917, -151.8610)
\qbezier(194.7917, -151.8610)(196.4509, -153.7822)(198.2541, -155.8292)
\qbezier( 77.3596, -47.4258)( 79.7229, -50.0216)( 81.8845, -52.4352)
\qbezier( 81.8845, -52.4352)( 84.0462, -54.8489)( 86.0203, -57.0962)
\qbezier( 86.0203, -57.0962)( 87.9944, -59.3434)( 89.7937, -61.4388)
\qbezier( 89.7937, -61.4388)( 91.5931, -63.5342)( 93.2293, -65.4914)
\qbezier( 93.2293, -65.4914)( 94.8656, -67.4485)( 96.3494, -69.2801)
\qbezier( 96.3494, -69.2801)( 97.8332, -71.1117)( 99.1742, -72.8296)
\qbezier( 99.1742, -72.8296)(100.5151, -74.5475)(101.7220, -76.1629)
\qbezier(101.7220, -76.1629)(102.9288, -77.7783)(104.0093, -79.3016)
\qbezier(104.0093, -79.3016)(105.0898, -80.8250)(106.0510, -82.2661)
\qbezier(106.0510, -82.2661)(107.0123, -83.7073)(107.8604, -85.0757)
\qbezier(107.8604, -85.0757)(108.7086, -86.4441)(109.4492, -87.7485)
\qbezier(109.4492, -87.7485)(110.1898, -89.0529)(110.8277, -90.3019)
\qbezier(110.8277, -90.3019)(111.4655, -91.5509)(112.0047, -92.7525)
\qbezier(112.0047, -92.7525)(112.5440, -93.9540)(112.9881, -95.1161)
\qbezier(112.9881, -95.1161)(113.4322, -96.2781)(113.7841, -97.4081)
\qbezier(113.7841, -97.4081)(114.1360, -98.5381)(114.3980, -99.6434)
\qbezier(114.3980, -99.6434)(114.6599, -100.7487)(114.8336, -101.8364)
\qbezier(114.8336, -101.8364)(115.0073, -102.9242)(115.0938, -104.0015)
\qbezier(115.0938, -104.0015)(115.1804, -105.0787)(115.1804, -106.1525)
\qbezier(151.0169, -106.1525)(151.0169, -107.1214)(151.0991, -108.0938)
\qbezier(151.0991, -108.0938)(151.1813, -109.0661)(151.3463, -110.0489)
\qbezier(151.3463, -110.0489)(151.5113, -111.0317)(151.7603, -112.0321)
\qbezier(151.7603, -112.0321)(152.0092, -113.0324)(152.3439, -114.0575)
\qbezier(152.3439, -114.0575)(152.6787, -115.0825)(153.1016, -116.1397)
\qbezier(153.1016, -116.1397)(153.5244, -117.1968)(154.0385, -118.2936)
\qbezier(154.0385, -118.2936)(154.5526, -119.3904)(155.1616, -120.5348)
\qbezier(155.1616, -120.5348)(155.7706, -121.6792)(156.4789, -122.8793)
\qbezier(156.4789, -122.8793)(157.1871, -124.0794)(157.9997, -125.3440)
\qbezier(157.9997, -125.3440)(158.8124, -126.6085)(159.7352, -127.9465)
\qbezier(159.7352, -127.9465)(160.6580, -129.2846)(161.6977, -130.7057)
\qbezier(161.6977, -130.7057)(162.7373, -132.1268)(163.9013, -133.6412)
\qbezier(163.9013, -133.6412)(165.0652, -135.1557)(166.3619, -136.7743)
\qbezier(166.3619, -136.7743)(167.6585, -138.3929)(169.0971, -140.1273)
\qbezier(169.0971, -140.1273)(170.5357, -141.8618)(172.1266, -143.7245)
\qbezier(172.1266, -143.7245)(173.7175, -145.5872)(175.4722, -147.5916)
\qbezier(175.4722, -147.5916)(177.2268, -149.5960)(179.1578, -151.7564)
\qbezier(179.1578, -151.7564)(181.0888, -153.9168)(183.2101, -156.2489)
\qbezier(183.2101, -156.2489)(185.3313, -158.5809)(187.6580, -161.1012)
\qbezier( 68.6132, -51.2038)( 70.9399, -53.7242)( 73.0611, -56.0562)
\qbezier( 73.0611, -56.0562)( 75.1824, -58.3882)( 77.1134, -60.5487)
\qbezier( 77.1134, -60.5487)( 79.0444, -62.7091)( 80.7990, -64.7135)
\qbezier( 80.7990, -64.7135)( 82.5537, -66.7179)( 84.1446, -68.5806)
\qbezier( 84.1446, -68.5806)( 85.7355, -70.4433)( 87.1741, -72.1777)
\qbezier( 87.1741, -72.1777)( 88.6127, -73.9122)( 89.9093, -75.5308)
\qbezier( 89.9093, -75.5308)( 91.2059, -77.1494)( 92.3699, -78.6638)
\qbezier( 92.3699, -78.6638)( 93.5339, -80.1783)( 94.5735, -81.5994)
\qbezier( 94.5735, -81.5994)( 95.6132, -83.0205)( 96.5360, -84.3585)
\qbezier( 96.5360, -84.3585)( 97.4588, -85.6966)( 98.2714, -86.9611)
\qbezier( 98.2714, -86.9611)( 99.0841, -88.2257)( 99.7923, -89.4258)
\qbezier( 99.7923, -89.4258)(100.5006, -90.6259)(101.1096, -91.7703)
\qbezier(101.1096, -91.7703)(101.7186, -92.9147)(102.2327, -94.0115)
\qbezier(102.2327, -94.0115)(102.7467, -95.1083)(103.1696, -96.1654)
\qbezier(103.1696, -96.1654)(103.5925, -97.2226)(103.9272, -98.2476)
\qbezier(103.9272, -98.2476)(104.2620, -99.2727)(104.5109, -100.2730)
\qbezier(104.5109, -100.2730)(104.7599, -101.2734)(104.9249, -102.2562)
\qbezier(104.9249, -102.2562)(105.0898, -103.2390)(105.1720, -104.2113)
\qbezier(105.1720, -104.2113)(105.2542, -105.1837)(105.2542, -106.1525)
\qbezier(141.0908, -106.1525)(141.0908, -107.2263)(141.1774, -108.3036)
\qbezier(141.1774, -108.3036)(141.2639, -109.3809)(141.4376, -110.4687)
\qbezier(141.4376, -110.4687)(141.6113, -111.5564)(141.8732, -112.6617)
\qbezier(141.8732, -112.6617)(142.1352, -113.7670)(142.4871, -114.8970)
\qbezier(142.4871, -114.8970)(142.8390, -116.0270)(143.2831, -117.1890)
\qbezier(143.2831, -117.1890)(143.7272, -118.3510)(144.2664, -119.5526)
\qbezier(144.2664, -119.5526)(144.8057, -120.7542)(145.4435, -122.0032)
\qbezier(145.4435, -122.0032)(146.0814, -123.2522)(146.8220, -124.5566)
\qbezier(146.8220, -124.5566)(147.5626, -125.8610)(148.4108, -127.2294)
\qbezier(148.4108, -127.2294)(149.2589, -128.5978)(150.2201, -130.0389)
\qbezier(150.2201, -130.0389)(151.1814, -131.4801)(152.2619, -133.0035)
\qbezier(152.2619, -133.0035)(153.3424, -134.5268)(154.5492, -136.1422)
\qbezier(154.5492, -136.1422)(155.7561, -137.7576)(157.0970, -139.4755)
\qbezier(157.0970, -139.4755)(158.4380, -141.1934)(159.9218, -143.0250)
\qbezier(159.9218, -143.0250)(161.4056, -144.8566)(163.0418, -146.8137)
\qbezier(163.0418, -146.8137)(164.6781, -148.7709)(166.4774, -150.8663)
\qbezier(166.4774, -150.8663)(168.2768, -152.9617)(170.2509, -155.2089)
\qbezier(170.2509, -155.2089)(172.2250, -157.4562)(174.3866, -159.8698)
\qbezier(174.3866, -159.8698)(176.5483, -162.2835)(178.9116, -164.8793)
\qbezier( 58.0170, -56.4759)( 59.8203, -58.5229)( 61.4795, -60.4441)
\qbezier( 61.4795, -60.4441)( 63.1387, -62.3654)( 64.6626, -64.1711)
\qbezier( 64.6626, -64.1711)( 66.1866, -65.9768)( 67.5832, -67.6765)
\qbezier( 67.5832, -67.6765)( 68.9799, -69.3762)( 70.2567, -70.9788)
\qbezier( 70.2567, -70.9788)( 71.5335, -72.5814)( 72.6972, -74.0955)
\qbezier( 72.6972, -74.0955)( 73.8608, -75.6095)( 74.9175, -77.0430)
\qbezier( 74.9175, -77.0430)( 75.9742, -78.4764)( 76.9294, -79.8369)
\qbezier( 76.9294, -79.8369)( 77.8847, -81.1973)( 78.7436, -82.4919)
\qbezier( 78.7436, -82.4919)( 79.6024, -83.7864)( 80.3695, -85.0220)
\qbezier( 80.3695, -85.0220)( 81.1365, -86.2575)( 81.8158, -87.4406)
\qbezier( 81.8158, -87.4406)( 82.4950, -88.6236)( 83.0900, -89.7604)
\qbezier( 83.0900, -89.7604)( 83.6851, -90.8972)( 84.1990, -91.9937)
\qbezier( 84.1990, -91.9937)( 84.7130, -93.0903)( 85.1486, -94.1523)
\qbezier( 85.1486, -94.1523)( 85.5842, -95.2144)( 85.9437, -96.2476)
\qbezier( 85.9437, -96.2476)( 86.3033, -97.2808)( 86.5887, -98.2906)
\qbezier( 86.5887, -98.2906)( 86.8741, -99.3004)( 87.0868, -100.2921)
\qbezier( 87.0868, -100.2921)( 87.2995, -101.2838)( 87.4407, -102.2627)
\qbezier( 87.4407, -102.2627)( 87.5819, -103.2416)( 87.6523, -104.2127)
\qbezier( 87.6523, -104.2127)( 87.7228, -105.1839)( 87.7228, -106.1525)
\put( 48.0508, -106.1525){\vector(1, 0){160.1695}}
\put(128.1356, -186.2373){\vector(0, 1){160.1695}}
\put(215.0847, -109.5847){\makebox(0,0)[]{\footnotesize{$\widehat{R}$}}}
\put(121.2712, -24.9237){\makebox(0,0)[]{\footnotesize{$\widehat{T}$}}}
\put(181.9068, -180.5169){\makebox(0,0)[]{\footnotesize{$\chi = \mbox{const}$}}}
\end{picture}
\vspace{-6mm} \newline
%
\vspace*{6mm} \newline
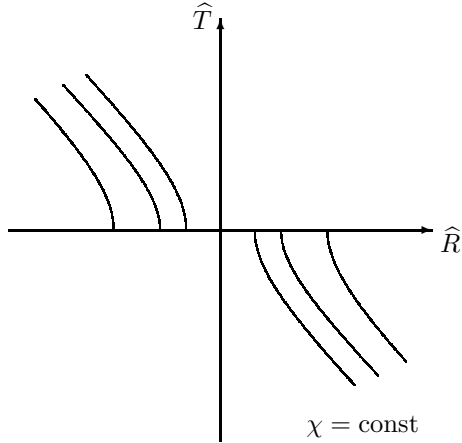
{\footnotesize \sf Figure 9. World lines of reference particles with
$\chi = \mbox{constant}$ (defining the Hubble flow) in a LIVE dominated
universe as observed from the CFS system defined by the transformation
\eqref{e_228}.}
\vspace{0mm} \newline
%
%
%
%
\vspace{5mm} \newline
{\bf 3. Penrose diagrams}
\vspace{3mm} \newline
By applying the rule in Appendix B of paper [1] for composing generating
functions one may obtain a deeper understanding of the conformal coordinate
transformations and find new ones. Let us first consider the
transformations to CFS coordinates of type III in universe models
with negative spatial curvature. Here the generating function in
II:(67) is obtained as a composition of the generating
functions in II:(55) and II:(71).
\itm Combining the generating functions \eqref{e_207} and II:(71)
we obtain the generating function
\begin{equation} \label{e_208}
f(x) = - \cot (x / 2)
\mbox{ .}
\end{equation}
and the transformation
\begin{equation} \label{e_209}
T = \frac{\sin \eta}{\cos \eta - \cos \chi}
\mbox{\hspace{2mm} , \hspace{3mm}}
R = \frac{\sin \chi}{\cos \chi - \cos \eta}
\mbox{ ,}
\end{equation}
In the same way as in section 2.4 this only represents a composition
of the antipodal transformation with the transformation \eqref{e_207},
and therefore nothing new is obtained physically.
\itm We have seen in section 3 of paper [1] that it is possible to
transform away spatial curvature. In the same way we obtain a more
general transformation from spaces with curvature $k_1$ to spaces with
curvature $k_2$ if the generating function $f$ satisfies the relation
\begin{equation} \label{e_132}
f'(u) f'(v) \hs{0.6mm} S_{k_1} \hs{-0.5mm} \left( \frl{u - v}{2} \right)^2
= S_{\hs{0.5mm} k_2} \hs{-0.5mm} \left( \frl{f(u) - f(v)}{2} \right)^2
\mbox{ .}
\end{equation}
%
Composition of generating functions can also be used to find coordinates
used in Penrose diagrams, as we will investigate below.
Our formalism can be utilized to give simple deductions of the
transformations that lead to the Penrose diagrams. Then equation
I:(29) for the CFS scale factor is modified to give a
corresponding conformal Einstein space (CES) scale factor, given as
\begin{equation} \label{e_298}
A(T,R) = \frl{a(\eta(T,R)) \hs{0.5mm} S_k(\chi(T,R))}{|\sin R \hs{0.7mm} |}
\mbox{ .}
\end{equation}
\vspace{-9mm} \newline
\itm By means of the composition rule in Appendix B of paper [1] we shall
first find a transformation from spaces with negative curvature to spaces
with positive curvature, corresponding to Einstein's static universe.
\itm By means of a composition of the generating function in II:(10)
and the inverse of the generating function in I:(28), we obtain the
generating function
\begin{equation} \label{e_305}
f(x) = - \frl{1}{T_i} e^{-x}
\mbox{ .}
\end{equation}
This represents a transformation from a space with $k = -1$ and CFS
coordinates of type I to a space with $k = 1$ and CFS coordinates of
type I. Equation I:(28) then implies that the transformed
coordinates $T$ and $R$ are given by
\begin{equation} \label{e_306}
\tan T = \frl{2 T_i e^{\eta} \cosh \chi}{1 - T_i^2 e^{2 \eta}}
\mbox{\hspace{2mm} , \hspace{3mm}}
\tan R = \frl{2 T_i e^{\eta} \sinh \chi}{1 + T_i^2 e^{2 \eta}}
\mbox{ .}
\end{equation}
The inverse transformation is
\begin{equation} \label{e_307}
T_i e^{\eta} = \sqrt{\frl{\cos R - \cos T}{\cos R + \cos T}}
\mbox{\hspace{2mm} , \hspace{3mm}}
\tanh \chi = \frl{\sin R}{\sin T}
\mbox{ .}
\end{equation}
Using equation \eqref{e_307} and the transformation I:(17) we
find that the line element I:(1) with $k = -1$ as expressed in
CES coordinates takes the form
\begin{equation} \label{e_310}
ds^2 = \frl{a(\eta(T,R))^2}{\cos^2 R - \cos^2 T} \hs{1.0mm}
(- dT^2 + dR^2 + \sin^2 R \hs{1.0mm} d \Omega^2)
= \frl{a(\eta(T,R))^2}{\cos^2 R - \cos^2 T} \hs{1.0mm} ds_E^2
\mbox{ .}
\end{equation}
where $ds_E^2$ is the line element of Einstein's static universe.
\itm In the case of the Milne universe the transformation to the
CES coordinates reduces to
\begin{equation} \label{e_308}
\tan T = \frl{t \cosh \chi}{1 - t^2}
\mbox{\hspace{2mm} , \hspace{3mm}}
\tan R = \frl{t \sinh \chi}{1 + t^2}
\mbox{ .}
\end{equation}
transforming the first quadrant $t > 0$ and $\chi > 0$ onto the
triangle $0 < R < \pi / 2$ and $R < T < \pi - R$, with inverse
transformation
\begin{equation} \label{e_309}
t = \sqrt{\frl{\cos R - \cos T}{\cos R + \cos T}}
\mbox{\hspace{2mm} , \hspace{3mm}}
\tanh \chi = \frl{\sin R}{\sin T}
\mbox{ .}
\end{equation}
The CES form of the line element for the Milne universe is
\begin{equation} \label{e_311}
ds^2 = \frl{1}{(\cos R + \cos T)^2} \hs{1.0mm} ds_E^2
\mbox{ .}
\end{equation}
\itm Taking a composition of the generating function in II:(55)
and the inverse of the generating function in \eqref{e_207},
we obtain the generating function
\begin{equation} \label{e_210}
f(x) = 2 \arctan (\tanh (x / 2))
\mbox{ .}
\end{equation}
This represents a transformation from a space with $k = -1$ and CFS
coordinates of type II to a space with $k = 1$ and CFS coordinates of
type I. Equation I:(28) then implies that the transformed
coordinates $T$ and $R$ are given by
\begin{equation} \label{e_40}
\tan T = \frl{\sinh \eta}{\cosh \chi}
\mbox{\hspace{2mm} , \hspace{3mm}}
\tan R = \frl{\sinh \chi}{\cosh \eta}
\mbox{ .}
\end{equation}
If one transforms from a universe model with dust and radiation which
begins at $\eta = 0$, this transformation maps the first quadrant
$\chi > 0$, $\eta > 0$ onto the triangle $0 < \widehat{T} < \pi / 2$
and $0 < \widehat{R} < \pi / 2 - \widehat{T}$. On the other hand, if
one transforms from a universe which is vacuum dominated and infinitely
old, the fourth quadrant $\chi > 0$, $\eta < 0$ is mapped onto the triangle
$-\pi / 2 < \widehat{T} < 0$ and $0 < \widehat{R} < \pi / 2 + \widehat{T}$.
The inverse transformation is
\begin{equation} \label{e_41}
\tanh \eta = \frl{\sin T}{\cos R}
\mbox{\hspace{2mm} , \hspace{3mm}}
\tanh \chi = \frl{\sin R}{\cos T}
\mbox{ .}
\end{equation}
Using equation \eqref{e_298} and the transformation \eqref{e_41} we
find that the line element I:(17) for a universe model with
$k = -1$ as expressed in CES coordinates takes the form
\begin{equation} \label{e_45}
ds^2 = \frl{a(\eta(T,R))^2}{\cos^2 T - \sin^2 R} \hs{1.0mm}
(- dT^2 + dR^2 + \sin^2 R \hs{1.0mm} d \Omega^2)
= \frl{a(\eta(T,R))^2}{\cos^2 T - \sin^2 R} \hs{1.0mm} ds_E^2
\mbox{ .}
\end{equation}
\itm We shall next find a transformation from flat spaces to spaces with
positive curvature, corresponding to Einstein's static universe.
Taking a composition of the generating function $f(x) = x$ for the
first type of CFS coordinates in flat space, and the inverse of the
generating function in \eqref{e_207}, we obtain the generating function
\begin{equation} \label{e_300}
f(x) = 2 \arctan (x)
\mbox{ ,}
\end{equation}
which is the inverse of the generating function given in formula
\eqref{e_207}. Hence the coordinate transformation is the inverse
of the one given in equation \eqref{e_49},
\begin{equation} \label{e_314}
\cot T = \frl{1 -
\left({\eta}^{2} - {\chi}^{2} \right)}{2 \eta}
\mbox{\hspace{2mm} , \hspace{3mm}}
\cot R = \frl{1 +
\left({\eta}^{2} - {\chi}^{2} \right)}{2 \chi}
\mbox{ .}
\end{equation}
with inverse transformation
\begin{equation} \label{e_301}
\eta = \frl{\sin T}{\cos T + \cos R}
\mbox{\hspace{2mm} , \hspace{3mm}}
\chi = \frl{\sin R}{\cos T + \cos R}
\end{equation}
Again using \eqref{e_298}, this time in combination with the
transformation \eqref{e_301}, we find the line element for flat
space in CES coordinates
\begin{equation} \label{e_303}
ds^2 = \frl{a(\eta(T,R))^2}{(\cos T + \cos R)^2} \hs{1.0mm}
(- dT^2 + dR^2 + \sin^2 R \hs{1.0mm} d \Omega^2)
= \frl{a(\eta(T,R))^2}{(\cos T + \cos R)^2} \hs{1.0mm} ds_E^2
\mbox{ .}
\end{equation}
\vspace{-9mm} \newline
\itm Note that the line element \eqref{e_311} can be immediately obtained
by putting $a = 1$ in equation \eqref{e_303}. This is a consequence of
the fact that the Milne universe is a coordinate transformation of the
interior of the future light cone of the point $(0,0)$ in the static
and spatially flat Minkowski spacetime.
\itm As a simple example we shall find the CES coordinates for the
Minkowski spacetime with line element
\begin{equation} \label{e_299}
ds^2 = -dt^2 + dr^2 + r^2 d \Omega^2
\mbox{ .}
\end{equation}
From equation \eqref{e_298} and the transformation \eqref{e_301}
it follows that the CES scale factor for the Minkowski spacetime
is [9]
\begin{equation} \label{e_302}
A(T,R) = \frl{1}{\cos T + \cos R}
\mbox{ .}
\end{equation}
\itm It may be noted that in the case of universe models with positive
spatial curvature, $\eta$ and $\chi$ are CES coordinates. For a de Sitter
universe with $k = 1$ one obtains by using equation \eqref{e_141},
the CES form of the line element
\begin{equation} \label{e_319}
ds^2 = \frl{1}{\widehat{H}_{\Lambda}^2 \cos^2 \eta \rule[-0mm]{0mm}{4.25mm}}
ds_E^2
\mbox{ ,}
\end{equation}
correcting an error in reference [10].
%
%
%
\vspace{5mm} \newline
{\bf 4. Some physical properties of universe models using CFS coordinates}
\vspace{3mm} \newline
The parametric time $\eta$ is widely used in cosmology because many
formulae become simpler using this than when expressed in termes of
cosmic time $t$. Important examples are the formulae I:(9) and I:(10)
for the radius of the particle horizon and the expressions II:(30),
II:(31) and \eqref{e_16}, \eqref{e_12} for the cosmic scale factor
in a curved universe with radiation and dust.
\itm One may wonder if there are any similar advantages in using the CFS
coordinates $T$ and $R$. According to G.Endean [5, 11-12] there are several.
He argues that the picture of the universe obtained with reference to the
CFS coordinates is more correct and relevant when interpreting observational
data than the usual picture referring to cosmic time $t$ (or the parametric
time $\eta$) and the standard radial coordinate $\chi$. The criterion for a
Big Bang is that $A(0,R) = 0$. This will be the case if $a$ converges to
$0$ faster than $B$ when $T \rightarrow 0$.
\itm Endean claims for example that there was no Big Bang because the
recession velocity vanishes at $t = 0$ (see equations II:(60)
and II:(75)). The recession velocity is the velocity
of a galaxy with $\chi = \mbox{constant}$ relative to the CFS system.
We prefer the following interpretation: The mentioned expressions tell
that the reference particles of the CFS system are initially at rest
relative to the galaxes. Then the CFS particles accelerate away from the
galaxes, which in no way implies that there was no Big Bang.
\itm Endean also claims that in a dust dominated universe with $k = 1$
there will not be a Big Crunch. The reason for this is the following.
At $\chi = 0$ the coordinate transformation \eqref{e_49} gives
\begin{equation} \label{e_127}
T = \tan \frl{\eta}{2}
\mbox{ .}
\end{equation}
Furthermore, the cosmic scale factor $a = \alpha (1 - \cos \eta)$ has a
maximum value $a_{\max} = 2 \alpha$ for $\eta = \pi$. This corresponds
to a finite cosmic time $t_1 = \pi \alpha$, but to an infinite CFS time.
Hence, in the CFS picture the closed dust dominated universe expands for
an infinitely long time, and a Big Crunch never happens.
\itm However, the CFS clocks are not standard clocks like the cosmic
clocks showing $t$. They are coordinate clocks going at the rate that is
adjusted to be in accordance with the first of equations \eqref{e_49}.
Compared to the standard clocks and clocks showing parametric time, the
CFS clocks go increasingly fast. Hence the non-vanishing of a Big Crunch
in the CFS picture is a coordinate effect. Further criticism of Endean's
interpretation of the CFS picture has been given by L.Querella [13].
\itm Let us consider a few examples in which the conformal time at
$\chi = 0$ can be very simply related to the cosmic time, namely
radiation dominated universes with negative, vanishing and positive
spatial curvature. With reference to the CFS systems given \linebreak
respectively in the transformations II:(49), II:(68) and
\eqref{e_49} we find for clocks at $\chi = 0$,
\begin{equation} \label{e_128}
T = \frl{t}{a}
\mbox{ .}
\end{equation}
In a radiation dominated universe [14,15]
\begin{equation} \label{e_129}
a = \sqrt{2 \beta t - k t^2}
\mbox{ .}
\end{equation}
Hence
\begin{equation} \label{e_130}
T = \sqrt{\frl{t}{2 \beta - kt}}
\mbox{ .}
\end{equation}
%
%
%
\vspace{5mm} \newline
{\bf 4.1. The CFS Hubble parameter of some universe models}
\vspace{3mm} \newline
The behaviour of the conformal space in different universe models
may be investigated by calculating the CFS Hubble parameter defined in
equation I:(48). We first consider negatively curved universe models
with respectively dust and radiation as described in CFS systems of
type I. Using equation II:(35) we obtain
\begin{equation} \label{e_253}
H_R = \frac{2 T_i T}{(\sqrt{T^2 - R^2} - T_i)^3}
\end{equation}
for a dust dominated universe, and by means of equation II:(36)
we get
\begin{equation} \label{e_254}
H_R = \frac{2 T_i^2 T}{(T^2 - R^2 - T_i^2)^2}
\end{equation}
for a radiation dominated universe. As shown in section 5 of paper [2]
there is continual creation at the boundary of the conformal space given
by $T^2 - R^2 = T_i^2$. Hence the CFS Hubble parameter for both of these
universe models approaches infinity at the boundary with continual
creation.
\itm The CFS Hubble parameter for a negatively curved radiation dominated
universe with conformal coordinates of type II and III is
\begin{equation} \label{e_293}
H_R = \mbox{\small sgn} (T) \hs{0.5mm}
\frl{(1 + \widehat{T}^2 - \widehat{R}^2)^2 - 4 \widehat{T}^2
(\widehat{T}^2 - \widehat{R}^2)}{4 \beta \widehat{T}^2}
\end{equation}
Correspondingly for a flat universe one obtains
\begin{equation} \label{e_294}
H_R = \frl{|T^2 - R^2| \hs{0.5mm} (3 T^2 + R^2)}{\beta T^2}
\end{equation}
For a universe with positive spatial curvature it is
\begin{equation} \label{e_295}
H_R = \mbox{\small sgn} (T) \hs{0.5mm}
\frl{[1 - (\widehat{T}^2 - \widehat{R}^2)]^2 - 4 \widehat{T}^2
(\widehat{T}^2 - \widehat{R}^2)}{4 \beta \widehat{T}^2}
\end{equation}
The Hubble parameter for a LIVE dominated universe model
with $k = -1$ and CFS coordinates of type I is
\begin{equation} \label{e_255}
H_R = \sqrt{\Omega_0} \hs{0.5mm} \widehat{H}_0 \hs{0.5mm} \frl{T}{T_f}
\end{equation}
In this case the conformal Hubble parameter in independent of the
position $R$.
The Hubble parameter for the same universe model but
with CFS coordinates of type II and III is
\begin{equation} \label{e_291}
H_R = - \widehat{H}_{\Lambda} \hs{0.9mm} \mbox{\small sgn} (T)
\end{equation}
where we have omitted the hat and the tilde on $T$.
This formula is also valid for a flat LIVE dominated universe.
The conformal space therefore expands for $T < 0$ and contracts
for $T > 0$. This behaviour is a result of two competing motions. The
Hubble flow expands exponentially. But as seen from Figure 12 in paper [2],
the conformal reference frame contracts relative to the Hubble flow. The
sign of conformal Hubble parameter shows that the expansion dominates for
$T < 0$ corresponding to the region $- \chi < \eta < 0$ in the
$(\eta,\chi)$-plane, while the contaction of the conformal system
dominates for $T > 0$ corresponding to the region $0 < \chi < - \eta$.
\itm From equations I:(48) and \eqref{e_216} it follows
that the CFS Hubble parameter for a LIVE dominated universe model
with positive spatial curvature in CFS coordinates of type I is
\begin{equation} \label{e_215}
H_R = \mbox{\small sgn} (R) \hs{0.5mm} \widehat{H}_{\Lambda} T
\mbox{ .}
\end{equation}
\vspace{-6mm} \newline
The corresponding Hubble parameter in CFS coordinates of type II and III
is
\begin{equation} \label{e_296}
H_R = \mbox{\small sgn} (R) \hs{0.5mm} \widehat{H}_{\Lambda}
\mbox{ .}
\end{equation}
\vspace{-8mm} \newline
%
%
%
%
\vspace{5mm} \newline
{\bf 4.2. The CFS age of some universe models}
\vspace{3mm} \newline
We shall now calculate the conformal age of some universe models
containing radiation only. They are of course not realistic universe
models, but we shall calculate their conformal age as an illustration.
\itm Consider first a negatively curved universe model. The parametric
age ${\eta}_0$ is determined from the normalization $a({\eta}_0) = 1$
in equation II:(30), which gives
\begin{equation} \label{e_283}
\sinh {\eta}_0 = 1 / \beta
\mbox{ .}
\end{equation}
The cosmic age is given in equation II:(31) with $\alpha = 0$.
Together with equation \eqref{e_283} this leads to
\begin{equation} \label{e_276}
t_0 = \sqrt{1 + {\beta}^2} - \beta
\mbox{ ,}
\end{equation}
where $\beta$ is given in equation II:(32) and has the value
$9.1 \cdot 10^{-3}$ for a radiation density equal to that of the
background radiation at the present time.
\itm We shall now calculate the present value of the time $T_0$ in a
CFS system of type I. It is given by equation II:(12) which shows
that $T_0$ is $R$-dependent and given by
\begin{equation} \label{e_269}
T_0 = \sqrt{R^2 + (1/4) \hs{0.5mm} (1 + \sqrt{1 + {\beta}^2})^2}
\mbox{ ,}
\end{equation}
The initial time of the universe model is
\begin{equation} \label{e_270}
T_i = \sqrt{R^2 + (1/4) \hs{0.5mm} {\beta}^2}
\mbox{ .}
\end{equation}
According to equation I:(5) the unit of time is
\begin{equation} \label{e_320}
l_0 / c = (H_0 \sqrt{1 - {\Omega}_{\gamma 0}})^{-1} = 13.9 \cdot 10^9 y
\mbox{ ,}
\end{equation}
where we have used the value of $H_0$ in reference [16], and that
$\Omega_{\gamma 0} = 6 \cdot 10^{-5}$. At $R = 0$ the dimensional
initial time is
\begin{equation} \label{e_267}
(l_0/c) \hs{0.5mm} T_i = \frl{l_0 \beta}{2c} = 63 \cdot 10^6 y
\mbox{ .}
\end{equation}
The CFS age of the universe is $T_0 - T_i$, where $T_0$ and $T_i$ are
given in equations \eqref{e_269} and \eqref{e_270}. Hence the conformal
age at $R = 0$ of this radiation dominated universe model is
\begin{equation} \label{e_268}
\frl{l_0}{c} \hs{0.5mm} (T_0 - T_i)
= \frl{l_0}{2c} \hs{0.5mm}
\{ (1 + \sqrt{1 + {\beta}^2}) - \beta \}
= 13.8 \cdot 10^9 y
\mbox{ .}
\end{equation}
Using equations \eqref{e_269} and II:(31) we find that the
conformal age and the cosmic age at $R = 0$ are related by
\begin{equation} \label{e_271}
T_0 - T_i = \frl{1}{2} (1 + t_0)
\mbox{ .}
\end{equation}
\itm Consider next a negatively curved universe model with CFS
coordinates of type II. The CFS age of this universe is
\begin{equation} \label{e_272}
\widehat{T}_0 = \sqrt{1 + {\beta}^2} - \sqrt{\widehat{R}^2 + {\beta}^2}
\mbox{ ,}
\end{equation}
where $0 < \widehat{T}_0 < 1$ and $0 < \widehat{R} < 1 - \widehat{T}_0$.
The conformal age and the cosmic age at $\widehat{R} = 0$ are related by
\begin{equation} \label{e_273}
\widehat{T}_0 =
\left\{ \begin{array}{lcl}
t_0          & \mbox{for} & 0 < t_0 < 1 \\
\vspace{-2.8mm} \\
\frl{1}{t_0} & \mbox{for} & t_0 > 1
\end{array} \right.
\mbox{ .}
\end{equation}
\itm We then consider a flat, radiation dominated universe.
The cosmic age of this universe is
\begin{equation} \label{e_282}
t_0 = \frl{1}{2 \beta}
\mbox{ ,}
\end{equation}
where $\beta$ is given in equation II:(77).
As described in a CFS coordinate system given in equation II:(72),
there are some strange coordinate effects. In the CFS coordinate system
the universe seems to be divided into two parts. The part $U_1$ outside
the future light cone of $(\eta,\chi) = (0,0)$ appears in the conformal
coordinate system with a positive time, coming from $\widehat{T} = 0$
and approaching infinity. The part $U_2$ inside the light cone appears
with a negative conformal time, coming from minus infinity and converging
towards $(\widehat{T},\widehat{R}) = (0,0)$.
\itm The conformal age of $U_1$ is
\begin{equation} \label{e_274}
T_0 = \frl{1}{2} (\sqrt{R^2 + {\beta}^2} - \beta)
\mbox{ .}
\end{equation}
For the other part $U_2$ one can only define the time left until a
Big Crunch of the conformal space, which is given by
\begin{equation} \label{e_275}
T_f - T_0 = \frl{1}{2} (\sqrt{R^2 + {\beta}^2} + \beta) - R
\mbox{ .}
\end{equation}
\itm We finally consider a positively curved, radiation dominated universe.
Its cosmic age is
\begin{equation} \label{e_279}
t_0 = \beta - \sqrt{{\beta}^2 - 1}
\mbox{ ,}
\end{equation}
where $\beta$ is given in equation \eqref{e_278}.
The radiation dominated universe is represented by the region
$\eta < \pi$ in Figure 2. This region has two parts I and II.
In part I the conformal time is positive, and the age is
\begin{equation} \label{e_284}
T_0 = \sqrt{R^2 + {\beta}^2} - \sqrt{{\beta}^2 - 1}
\mbox{ .}
\end{equation}
On the other hand, in part II the conformal time is negative, and the
conformal space disappears at $\widetilde{T} = 0$.
In this case the conformal time left until the space disappears is
\begin{equation} \label{e_280}
- T_0 = \sqrt{R^2 + {\beta}^2} + \sqrt{{\beta}^2 - 1}
\mbox{ .}
\end{equation}
%
%
%
\vspace{5mm} \newline
{\bf 4.3. Recession velocity and cosmic redshift in CFS space}
\vspace{3mm} \newline
We have considered several types of conformally flat spacetime
coordinate systems. One series has $a = b = d = 0$ in equation
I:(33). Then the generating function reduces to
\begin{equation} \label{e_246}
f(x) =  c / I_k (x/2)
\mbox{ ,}
\end{equation}
and the general expression I:(46) for
the recession velocity takes the form
\begin{equation} \label{e_247}
V = \frl{2kRT}{c^2 + k \hs{0.3mm} (R^2 + T^2)}
\mbox{ .}
\end{equation}
\itm The corresponding formula for the Doppler shift factor given in
equation I:(50) is
\begin{equation} \label{e_261}
D(T,R) = \left\{ \frl{c^2 + \hs{0.3mm} k \hs{0.3mm}
( \hs{0.3mm} T + R \hs{0.5mm} )^2}
{c^2 + \hs{0.3mm} k \hs{0.3mm}
( \hs{0.3mm} T - R \hs{0.5mm} )^2} \right\}^{1/2}
\mbox{ .}
\end{equation}
\itm For $k = 0$ equation I:(50) gives
\begin{equation} \label{e_262}
D(T,R) = \frl{b \hs{0.3mm} ( \hs{0.3mm} T + R - d \hs{0.5mm} ) - c}
{b \hs{0.3mm} ( \hs{0.3mm} T - R - d \hs{0.5mm} ) - c}
\mbox{ .}
\end{equation}
We shall now find an expression for the redshift $z$ of an object at
$R$ emitting light at a time $T$ as observed by an observer at $R = 0$
at a time $T_0$. Since light has a contant velocity $1$ in the CFS
coordinates, this means that $R = T_0 - T$.
In the case of the Einstein de Sitter universe, i.e. a flat dust dominated
universe, the constants in equation \eqref{e_262} have the values $b = 2$,
$c = 2$ and $d = -1$, giving
\begin{equation} \label{e_263}
D(T,R) = \frl{T_0}{2 T - T_0}
\mbox{ .}
\end{equation}
Using this together with equation I:(51) we find for the CFS
redshift $z$,
\begin{equation} \label{e_264}
1 + z = \left( \frl{2 T - T_0}{T} \right)^2
\mbox{ .}
\end{equation}
The corresponding expression for the redshift i a flat LIVE dominated
universe, i.e. the de Sitter universe is
\begin{equation} \label{e_265}
1 + z = \frl{T}{2 T - T_0}
\mbox{ .}
\end{equation}
\itm Using equation I:(51) we find that the cosmic redshift $z$
in a flat universe with $f(x) = - 1/x$ is given by
\begin{equation} \label{e_176}
1 + z = \frl{T + R}{T - R} \hs{1.5mm} \frl{A(T_0,0)}{A(T,R)}
\mbox{\hspace{2mm} , \hspace{2mm}}
R = T_0 - T
\end{equation}
where $T$ is the emission time and $T_0$ the time of observation.
\itm The redshift $z$ in a universe with $k = -1$,
$f(x) = \tanh (x/2)$
\begin{equation} \label{e_266}
1 + z = \sqrt{\frl{1 - (T + R)^2}{1 - (T - R)^2}} \hs{1.5mm}
\frl{A(T_0,0)}{A(T,R)}
\mbox{ .}
\end{equation}
and in a universe with $k = 1$,
$f(x) = \tan (x/2)$ are given by the expression
\begin{equation} \label{e_177}
1 + z = \sqrt{\frl{1 + (T + R)^2}{1 + (T - R)^2}} \hs{1.5mm}
\frl{A(T_0,0)}{A(T,R)}
\mbox{ .}
\end{equation}
%
%
%
%
\vspace{5mm} \newline
{\bf 5. Conclusion}
\vspace{3mm} \newline
FRW universe models with positive spatial curvature have been described
in CFS coordinates. As in models with negative and vanishing curvature,
described in paper II in this series, the motion of the free particles
defining the Hubble flow with cosmic time, is utterly strange in these
coordinates.
\itm We have seen that in order to describe a univers model with positive
spatial curvature using CFS coordinates, the universe must be covered by
two coordinate domains, which we have called I and II in Figure 2. Let
us first describe the evolution of the CFS space in the domain I.
Physically the FRW universe with radiation and dust and with positive
curvature comes from a Big Bang singularity with vanishing spatial
extension. The corresponding coordinate range of the coordinate
$\chi$ which is comoving with the Hubble flow is
$< \hs{-0.5mm} 0, \pi \hs{-0.5mm} >$
at an arbitrary time. The CFS coordinate space appears as a Big Bang
happening everywhere at $T = 0$, and the universe expands. In the
infinitely far future a new CFS coordinate system
$(\widetilde{T},\widetilde{R})$ must be introduced, and the clocks
are reset from plus to minus infinity. In this coordinate system
the CFS space first contracts with a decreasing velocity which
stops at $\widetilde{T} = 0$, and then expands. Hence the evolution
of the CFS coordinate space gives the impression of a space that
accelerates outwards. This is however an illusion due to the motion
of the CFS reference frame.
\itm Thereafter a spherical hole in which space, dust and radiation
vanish, appears at $\widetilde{R} = 0$ at the point of time
\begin{equation} \label{e_600}
\widetilde{T} \cdot \frl{l_0}{c}
= \frl{\alpha}{\beta} \cdot \frl{l_0}{c}
= \frl{\Omega_{m 0}}{2 \sqrt{\Omega_{\gamma 0}} (\Omega_{0} - 1) H_0}
\mbox{ ,}
\end{equation}
where we have reintroduced the time unit
$l_0 / c = H_0^{-1} (\Omega_0 - 1)^{-1/2}$
defined in section 2 of reference [1].
This corresponds to a parametric time where the FRW space contracts.
The CFS bubble expands with superluminal velocity as illustrated
in Figure 6.
In this case there is continual annihilation of space, matter and
energy. Particles with $\chi > {\chi}_3$ hit the bubble and vanish,
while particles with $\chi < {\chi}_3$ avoid the bubble. They
disappear at an infinitely far future, only to reappear at an infinitely
remote past, now approaching the observer.
\itm This bubble does not appear if there is only dust or only radiation
in the universe. In a universe with dust only, $\beta = 0$, we have that
${\chi}_3 = \pi$. The FRW universe only exists for $0 \le \chi \le \pi$.
Hence if ${\chi}_3 = \pi$, there is no Hubble observer hitting the bubble.
This means that no bubble appears. In a universe with positive spatial
curvature containing dust only the universe ends at a parametric time
${\eta}_3 = 2 \pi$. In this case the hyperbola $\eta = {\eta}_3$ in
Figure 6, representing the surface of the bubble in which space vanishes,
is replaced by the negative $\widetilde{R}$-axis and the positive
$R$-axis. This means that the whole CFS space suddenly vanishes
at the point of time $T = 0$, corresponding to a Big Crunch.
A universe with radiation only will evolve in a similar manner allthough
${\eta}_3 = \pi$ in this case. Hence the continual annihilation of the
CFS space in a universe with dust and radiation is replaced by a
Big Crunch in universes with dust or radiation only.
\itm In a LIVE dominated universe there is a similar behaviour.
Considering first the region $0 < \chi < \pi / 2$ there is initially a
hole which is arbitrarily large in the infinitely remote past.
It decreases in size and vanishes at a
point of time $T = -1$. Hence for $T < -1$ there is continual creation of
CFS space and vacuum energy at the boundary of the hole. At $T = 1$ the
hole reappears. It expands, and for $T > 1$ there is continual
annihilation of space and energy. In the region $\pi / 2 < \chi < \pi$
the behaviour is more complicated. A bubble of conformal space appears
at $\widetilde{T} = 1$ and expands. In the limit as $\widetilde{T}$
approaches infinity this region of the universe evolves from the region II
into the region I in Figur 7. The clocks are then reset to come from minus
infinity, and the conformal radial coordinate changes from negative to
positive values. The CFS coordinate space of the universe now contracts,
but it accelerates outwards so that the contraction is replaced by an
expansion at $T = 0$. Then, as $T$ approaches infinity, the region
$\pi / 2 < \chi < \pi$ of the universe evolves from the region I into
region II. Finally this region of the universe contracts and is
continually annihilated, vanishing at $\widetilde{T} = -1$.
\itm In part II of this series we have considered transformations from a
description of FRW universe models with negative and vanishing spatial
curvature as described with cosmic time and standard radial coordinates
to a description in terms of CFS coordinates. In section 3 of the
present paper we have shown that by composing these transformations
with the inverse of transformations described in section 2 we obtain
a form of the line element with a scale factor times Einstein's
static universe which has positive spatial curvature. In this way
we obtain Penrose diagrams where infinitely remote regions in
spacetime is mapped onto a finite region.
\itm Finally we have deduced formulae for some physical properties of
the FRW universe models using CFS coordinates. In particular we have
found formulae for the CFS Hubble parameter, the age, the recession
velocity and the cosmic redshift of some universe models.
\itm Maybe the most important difference between the standard coordinates
and the CFS coordinates is that the standard coordinates are comoving
with a cosmic reference frame consisting of free particles defining
the Hubble flow, while the CFS coordinates are comoving with an
accelerated reference frame. The reference particles of the cosmic frame
have a recession velocity $V$ relative to the CFS system. This implies
that the  CFS simultaneity is different from the cosmic simultaneity.
Therefore the CFS space is different from the cosmic space.
\itm We have shown in the present series of papers that this difference
makes the CFS description of the universe radically different from the
description in terms of cosmic time and space. For instance, in some
universe models the Big Bang and the Big Crunch in the FRW description
are replaced by continual creation and annihilation. Both descriptions
are in accordance with the observations of our universe and concern
the same reality, but they refer to time coordinates with different
simultaneities.
\vspace{5mm} \newline
{\bf References}
%
\begin{enumerate}
\item \O .Gr\o n and S.Johannesen, \textit{FRW Universe Models in
Conformally Flat Spacetime Coordinates. I: General Formalism},
Eur.Phys.J.Plus \textbf{126}, 28 (2011).

\item \O .Gr\o n and S.Johannesen, \textit{FRW Universe Models in
Conformally Flat Spacetime Coordinates. II: Universe models with
negative and vanishing spatial curvature},
Eur.Phys.J.Plus \textbf{126}, 29 (2011).

\item G.U.Varieschi, \textit{A Kinematical Approach to Conformal Cosmology},
Gen.Rel.Grav. \textbf{42}, 929 - 974 (2010).

\item  M.Ibison, \textit{On the conformal forms of the
Robertson-Walker metric}, J.Math.Phys. \textbf{48},
122501-1 -- 122501-23 (2007).

\item G.Endean, \textit{Cosmology in Conformally Flat Spacetime}, The
Astrophysical Journal \textbf{479}, 40 - 45 (1997).

\item E.Eriksen and \O .Gr\o n, {\it The De Sitter Universe Models},
Int.J.Mod.Phys. {\bf D4}, 115 - 159 (1995).

\item  A.Lasenby and C.Doran, {\it Closed Universes, de Sitter Space and
inflation}. Phys.Rev. \textbf{D71}, 063502 (2005).

\item V.F.Mukhanov, \textit{Physical foundations of cosmology},
Cambridge University Press, (2005).

\item S.M.Carroll, {\it Spacetime and Geometry}, Addison Wesley, (2004),
Appendix H.

\item S.W.Hawking and G.F.R.Ellis, {\it The large scale structure of
space-time}, Cambridge University Press, (1973), equation (5.8).

\item G.Endean, \textit{Redshift and the Hubble Constant in Conformally
Flat Spacetime}, The Astrophysical Journal \textbf{434}, 397 - 401 (1994).

\item G.Endean, \textit{Resolution of Cosmological age and redshift-distance
difficulties}, Mon.Not. R.Astron.Soc. \textbf{277}, 627 - 629 (1995).

\item L.Querella, \textit{Kinematic Cosmology in Conformally Flat Spacetime},
The Astrophysical Journal \textbf{508}, 129 - 131 (1998).

\item \O .Gr\o n and S.Hervik, Einstein's General Theory of Relativity,
Springer, (2007), ch.11.

\item \O .Gr\o n, Lecture Notes on the General Theory of
Relativity, Springer, 2009, p 208.

\item S.H.Suyu et.al., {\it Dissecting the gravitational lens B1608+656.
II. Precision measurements of the Hubble constant, spatial curvature,
and the dark energy energy equation of state}, arXiv:0910.2773.

\end{enumerate}

\end{document}